\documentclass[aps,twocolumn,longbibliography] {revtex4-2}
\usepackage{amsmath,amssymb,graphicx,amsfonts}

\usepackage{times}
\usepackage[colorlinks=true,linkcolor=blue,urlcolor=blue,citecolor=blue,breaklinks=true]{hyperref}

\usepackage{comment}

\usepackage[utf8]{inputenc}

\def\da{\dagger}

\def\eq#1{Eq.~(\ref{#1})}
\def\da{\dagger}

\date\today
\begin{document}

\title{Bloch oscillations of a mobile impurity in a one-dimensional Bose gas}

\author{Saptarshi Majumdar}
\author{Aleksandra Petkovi\'{c}}
\affiliation{Universit\'{e} de Toulouse, CNRS, Laboratoire de Physique Th\'{e}orique, Toulouse, France}

\begin{abstract}

We study the motion of an impurity under the action of a constant force through a one-dimensional system of weakly-interacting bosons. The interplay of the impurity-boson interaction, the boson-boson interaction, and the driving force gives rise to a rich dynamics. We focus on the influence of a finite external force. Under these far-from-equilibrium conditions, we show that in a wide range of forces, one part of the momentum transferred to the system is periodically channeled into the Bose gas through the emission of dispersive density shock waves, solitons, density waves and the creation of additional phase gradients. As a result, the impurity velocity does not increase indefinitely, but periodically oscillates in time around the drift velocity. We uncover and characterize different dynamical regimes in a wide range of  the impurity-boson coupling, the impurity mass and the external force. At a sufficiently large force, the Bloch oscillations cease and the impurity exhibits an unlimited acceleration.

\end{abstract}

\maketitle

\section{Introduction}

An electron in a crystal lattice exposed to a uniform weak static electric field performs a periodic motion around a fixed position in space, known as the Bloch oscillations \cite{Bloch,Zener}. The time period is determined by the lattice constant $a$ as $T_B=2\pi \hbar/(a F)$, where $F$ is the force acting on the electron and $\hbar$ is the reduced Planck's constant. The Bloch oscillations have not been observed in conventional solids due to the lattice defects causing the coherence time to be much shorter than the Bloch period. However, the Bloch oscillations of electrons have been observed in semiconductor superlattices \cite{PhysRevLett.70.3319}, where the much bigger lattice constant yields a much shorter Bloch period. 
Thanks to the remarkable experimental control achieved in the field of cold atomic gases, the Bloch oscillations of atoms in periodic optical potentials have been observed \cite{PhysRevLett.76.4508,PhysRevLett.76.4512}.

Even more interesting are the Bloch oscillations of a distinguishable particle (an impurity) under the influence of a small external force in one-dimensional (1d) quantum systems in the absence of a periodic potential \cite{gangardt2009bloch}. The phenomenon is closely related to the peculiar property of 1d quantum liquids in the presence of an impurity, where the lowest-energy state is a periodic function of momentum, with a period $2\pi \hbar n_0$ \cite{lamacraft2009dispersion}. Here $n_0$ denotes the mean particle density of the host liquid. For an infinitesimal force,  the Bloch period reads as $T=2\pi \hbar n_0/F$, where $1/n_0$ plays the role of the lattice constant. Despite the absence of a long-range crystalline order in 1d, the experimental evidence of the Bloch oscillations of an impurity atom has been observed in a strongly-interacting system of bosons \cite{Meinert945}. 
Recently, the Bloch oscillations of a large number of distinguishable weakly-interacting particles forming a soliton have been predicted \cite{ThBlochSolitons,PhysRevLett.130.220403} and observed \cite{BlochSolitons} in a 1d Bose gas in the absence of a lattice potential.

The dynamics of a single impurity under the action of a constant force in 1d quantum liquids without any periodic potential has attracted considerable attention \cite{gangardt2009bloch,AnnalsKamenev,Schecter_2016,PhysRevE.90.032132,Meinert945}. An important progress was made in Refs.~\cite{gangardt2009bloch,AnnalsKamenev}, where the depletion cloud of the host particles around the impurity is assumed to be in a local equilibrium, while the Luttinger liquid theory is used to describe the non--equilibrium host liquid. This approach is limited to very small external forces. It takes into account only the low-energy excitations, i.e., the phonons with a liner dispersion relation. 
 In Ref.~\cite{PhysRevE.90.032132}, by employing the Boltzmann equation, the authors studied an impurity accelerated by a small force and weakly coupled to a Tonks-Girardeau gas. 
The conclusions from the above mentioned works disagree, and conditions under which the Bloch oscillations are realized are debated \cite{PhysRevE.92.016101,PhysRevE.92.016102}.
The system of both strongly-interacting impurity and bosons has been studied experimentally and numerically for finite forces in Ref.~\cite{Meinert945}, for an impurity of the same mass as the bosons.

In this work, we study the dynamics of a mobile impurity driven by an external force through a system of weakly-interacting 1d bosons. We take advantage of a weak repulsion between the bosons and solve the time-dependent mean-field equation of motion for bosons in the reference frame co-moving with the impurity. We study the impurity motion in a wide range of the driving force, the impurity coupling strength, and its mass. Our main motivation is to understand the impurity dynamics under the action of a finite driving force, as well as the accompanying dynamical response of the bosons. 
In order to gain an understanding of this far-from-equilibrium situation, we consider the time evolution of the density and the phase of the background bosons. We demonstrate that in a wide range of forces, one part of the momentum transferred to the system is periodically channeled into the Bose gas through the emission of dispersive density shock waves, solitons, density waves and the creation of additional phase gradients. As a result, the impurity experiences a friction force and in a wide range of forces its velocity does not increase indefinitely, but periodically oscillates in time around the drift velocity. 
We address the following questions. How does the impurity dynamics behave as a function of the external force, as well as  the strength of the impurity-boson and the boson-boson interactions? What kind of excitations are emitted and at what characteristic moments of the impurity dynamics? What is the range of external forces where the oscillations persist? How different is the dynamics in the case of a light and a heavy impurity?

The paper is organized as follows. Sec.~\ref{sec:model} introduces the model. In Sec.~\ref{sec:Stationary}, we discuss a finite-momentum ground state of the system, in the absence of a driving force.
Sec.~\ref{sec:general} investigates the dynamics of the impurity subject to an external force, as well as the reaction of the hosting bosons for a moderate impurity-boson coupling. We study the impurity drift velocity, the amplitude of the impurity velocity oscillations, as well as their shape and the time period as a function of the driving force. We also investigate the density of the background bosons and their phase as a function of time and space, and characterize the emitted excitations. Although the system is far from its ground state, the drifted Bloch oscillations persist in a wide interval of the driving force.
Sec.~\ref{sec:strong} studies the case of a strongly-coupled impurity. Sec.~\ref{Sec:dependence} investigates the influence of the impurity-boson coupling, the impurity mass and the boson-boson interaction on the dynamics of the system. The paper concludes with Sec.~\ref{sec:conclusions} that presents a summary of the main findings and their discussion.

\section{Model\label{sec:model}}

We study a single mobile impurity immersed into a system of 1d bosons with contact repulsion at zero temperature. The impurity experiences a constant driving force $F$. The Hamiltonian of the system is given by
\begin{align}
	\hat{H}=\frac{\hat{P}^2}{2M}+\hat{H}_b+G \hat{\Psi}^\da(\hat{X})\hat{\Psi}(\hat{X})-F \hat{X}.
	\label{eq:H}
\end{align}
The impurity momentum and position operators are denoted by $\hat{P}$ and $\hat X$, respectively. 
The Bose gas is described by the Hamiltonian $\hat H_b$, that takes the form
\begin{align}
	\hat{H}_b=\int \mathrm{d}x\left[-\hat\Psi^\da(x)\dfrac{\hbar^2\partial_x^2}{2m}\hat\Psi(x)+\frac{g}{2}\hat\Psi^\da(x)\hat\Psi^\da(x)\hat\Psi(x)\hat\Psi(x)\right]
\end{align}
Here, $\hat\Psi^\da(x)$ and $\hat\Psi(x)$ are the bosonic single-particle operators, which satisfy the commutation relation $[ \hat\Psi(x) , \hat\Psi^\da(x')]=\delta(x-x')$. The strength of the repulsion between the bosons is $g$, that can be characterized by the dimensionless parameter $\gamma=m g/\hbar^2 n_0$. Here, the mass of bosons is denoted by $m$, and their mean density by $n_0$. The impurity has a mass $M$ and locally couples to the density of bosons, with a coupling constant $G$. The impurity-boson interaction is assumed to be repulsive, $G>0$. 

In order to treat the problem, we proceed in two steps. First, we apply a time-dependent unitary transformation $\hat{U}_1(t)=e^{i F \hat{X} t/\hbar}$ as
$
\hat{H}_1(t)=\hat{U}_1^\da \hat{H}\hat{U}_1-i\hbar \hat{U}_1^\da \partial_t \hat{U}_1.
$
This transformation allows for the periodic boundary conditions for the Hamiltonian $\hat{H}_1(t)$, while the impurity momentum becomes $\hat{U}_1^\da \hat{P}\hat{U}_1=\hat{P}+Ft$.
Next, we apply the Lee-Low-Pines transformation \cite{LeeLowPines}, that acts as $\hat{\mathcal{H}}=\hat{U}_2^\da \hat{H}_1\hat{U}_2$ where $\hat{U}_2=e^{-i \hat{X} \hat{p}_b/\hbar}$. Here, the momentum of the Bose gas is $\hat{p}_b=-i\hbar\int \mathrm{d}x  \hat\Psi^\da(x)\partial_x \hat\Psi(x)$.
The impurity momentum transformes as $\hat{U}_2^\da \hat{P}\hat{U}_2=\hat{P}-\hat{p}_b$. 
The bosons are translated by the impurity position $\hat{X}$, $\hat{U}_2^\da \hat{\Psi}(x)\hat{U}_2=\hat{\Psi}(x-\hat{X})$, and in the new referent system the impurity is situated at the origin. Thus, the Hamiltonian $\hat{\mathcal{H}}$ does not depend on $\hat{X}$, and as a result, $\hat{P}$ is conserved in time and can be replaced by a number $p$. In the new reference frame, the total momentum of the system is
$
\hat{U}_2^\da \hat{U}_1^\da (\hat{P}+\hat{p}_b)\hat{U}_1 \hat{U}_2=\hat{P}+Ft=p+Ft.
$
The final Hamiltonian reads as 
\begin{align}\label{eq: HLee-Low}
	\hat{\mathcal{H}}(t)=\frac{(p+F t-\hat{p}_b)^2}{2M}+\hat{H}_b+G \hat{\Psi}^\da(0)\hat{\Psi}(0).
\end{align}

In the following, we are interested in weakly-interacting bosons, $\gamma\ll 1$. Then, the bosonic single-particle field operator can be written as \cite{pitaevskii_bose-einstein_2003,sykes_drag_2009,CasimirNewJPhys}
	$
		\hat\Psi(x,t)=\Psi_0(x,t)+\gamma^{1/4}\hat\Psi_1(x,t)+\ldots, \label{eq:wf_exp}
	$
where the field $\Psi_0(x,t)$ describes the condensate wave function in the absence of  fluctuations, while the higher order contributions account for the effects of quantum and thermal fluctuations. The mean-field equation of motion reads as \cite{PhysRevA.100.013619,atoms10010003,Will_2023}
\begin{align}
i\hbar \partial_t{{\Psi_0}(x,t)}=\Bigg[&-\frac{\hbar^2}{2}\left( \frac{1}{m}+\frac{1}{M}\right)\partial_x^2+g |{\Psi_0}(x,t)|^2\notag\\&+G\delta(x)+i\hbar V(t)\partial_x\Bigg] {\Psi_0}(x,t).
\label{eq:mean-field1}
\end{align}
Here, the impurity velocity takes the form 
\begin{align}\label{eq:Vimp}
V(t)=\frac{p+F t}{M}+i\frac{\hbar}{M}\int \mathrm{d}x  \Psi_0^*(x,t)\partial_x \Psi_0(x,t).
\end{align}
Note that the mean-field approach is not reliable for a light weakly-coupled impurity, where the effects of quantum fluctuations become more relevant.

\section{Finite-momentum ground state in the absence of the force}\label{sec:Stationary}

\begin{figure}
	\centering
	\includegraphics[width=\columnwidth]{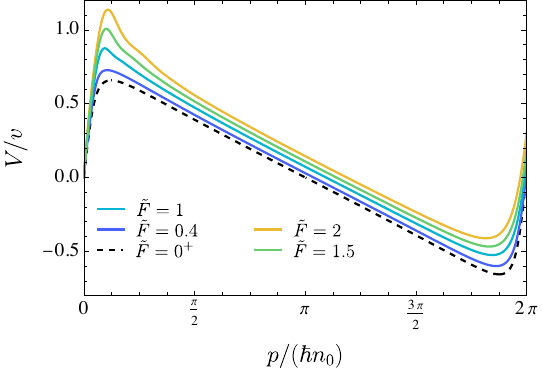}
	\caption{Impurity velocity as a function of the system momentum $p=F t$ for $0\leq t\leq  2\pi \hbar n_0/F$ is shown for different values of $F$. In the case of an  infinitesimal force, the velocity of the finite-momentum ground state is realized. It is studied in Sec~\ref{sec:Stationary}. The system parameters are $M=3m,\gamma=1/400$ and $\tilde{G}=G/\hbar v=0.5$. Here, $\tilde{F}=F \xi/g n_0$.}
	\label{fig1}
\end{figure}

In this section, we consider a stationary solution of \eq{eq:mean-field1} with the lowest-energy, for $F=0$ and a finite system momentum $p$. In the stationary state, the impurity velocity is constant, $V(t)=V$. We consider periodic boundary conditions. The stationary solution can be evaluated analytically \cite{Hakim}. Here we discuss some properties of the stationary state that are relevant for the subsequent sections.  For more details see Ref.~\cite{Quench} that reviews its properties.

The impurity modifies the boson density over the length scale $\xi \sqrt{1+m/M}/b$ around its position. Here the healing length is $\xi=\hbar/\sqrt{m g n_0}$, and the sound velocity is $v=\sqrt{g n_0/m}$. We also define $a={V}/{v\sqrt{1+m/M}}$ and $b=\sqrt{1-a^2}$. The boson density in the reference frame comoving with the impurity  is given by
\begin{align}
n(x)=n_0\left[a^2+b^2\tanh^2{\left(\frac{b|x|}{\xi\sqrt{1+m/M}} +x_0\right)}\right].
\label{densityStationary}
\end{align}
Here, the parameter $x_0$ satisfies
$G(a^2+b^2 \tanh^2{x_0})=\hbar v b^3 \sqrt{1+m/M} \tanh{x_0}\; \textrm{sech}^2{x_0} $. This equation admits solutions only for the impurity velocity smaller than the critical velocity, $|V|\leq v_c$. The critical velocity $v_c$ obeys the following equation \cite{Hakim}
\begin{align}\label{eq:vc}
	\frac{G}{\hbar v}\frac{1}{\sqrt{1+m/M}}=\frac{\sqrt{1-20
			\tilde{v}_c^2-8 \tilde{v}_c^4+\left(1+8 \tilde{v}_c^2\right)^{3/2}}}{2 \sqrt{2}\tilde{v}_c}.
\end{align}
Here, we have introduced $\tilde{v}_c=v_c/v \sqrt{1+m/M}$. 

The impurity also affects the phase $\Theta$ of the stationary condensate wave function $\Psi_0(x,t)=\sqrt{n(x)}\exp{\left[i \Theta(x)-i\mu t/\hbar\right]}$. The phase drop $\theta$ across the impurity position is given by
\begin{align}
\theta=&2 \arctan \left(\frac{b}{a} \right) -2\arctan\left(\frac{b }{a}\tanh{x_0}\right).
\label{eq:theta}
\end{align} 
As a result, the periodic boundary conditions impose a contribution $\theta x/L$ in the phase of the condensate wave function. Here the system length is denoted by $L$. 

The total momentum of the system takes the form
\begin{align}
p=&M{V}- N m V/(1+m/M) +\hbar n_0 \theta + 2 k\pi\hbar n_0\label{eq:momentum},
\end{align}
for the momentum in the interval $\pi \hbar n_0 (2k-1) <p\leq\pi\hbar n_0(2 k+1)$. Here $k$ is an integer. The number of expelled particles is $N=\int \mathrm{d} x [n_0-n(x)]$, and reads as
\begin{align}
N=&\frac{2b}{\sqrt{\gamma}}\sqrt{1+m/M}\left(1 - \tanh{x_0} \right) \label{eq:Nexpelled}.
\end{align}
There are four contributions in  \eq{eq:momentum}. The first and the second contribution in \eq{eq:momentum} are the momentum of the impurity and the hole around the impurity, respectively. The third term in \eq{eq:momentum} originates from the constant phase derivative in the remaining part of the system, that compensates the phase drop at the impurity. Next, we explain the fourth contribution. For $k=0$, the momentum satisfies $-\pi \hbar n_0<p\leq \pi \hbar n_0$ and the fourth term is zero. If the momentum of each particle is increased for $2k\pi\hbar/L$, the phase of the condensate wave function acquires an additional contribution $2k\pi x/L$, while the total momentum of the system is increased by $2k\pi\hbar n_0$, giving the last term in \eq{eq:momentum}. Note that this contribution is carried by supercurrents, and thus the energy remains unchanged. As a result, the ground-state energy of the system $E(p)$ is a periodic function of  the momentum $p$, with a period $2\pi\hbar n_0$ \cite{lamacraft2009dispersion,kamenev2009dynamics,AnnalsKamenev}. The impurity velocity is $V=\partial_{p} E(p)$, and thus it is also a periodic function of $p$. The impurity velocity as a function of $p$ is shown in Fig.~\ref{fig1} by the black dashed curve.

The boson density depletion at the impurity position $n_0-n(0)$ follows from \eq{densityStationary}. It is an increasing function of the total system momentum $p$ in the interval $0\leq p\leq \pi \hbar n_0$, and reaches its
maximal value $n_0$. Note that $n(0)$ depends only on the absolute value of the impurity velocity $|V|$. Thus, the boson density depletion is a decreasing  function of the momentum $p$ for $\pi \hbar n_0\leq p\leq 2\pi\hbar n_0$.
\begin{figure*}
	\centering
	\includegraphics[width=0.49\linewidth]{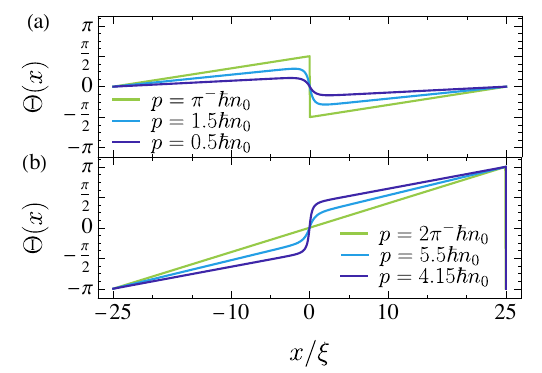}
	\includegraphics[width=0.49\linewidth]{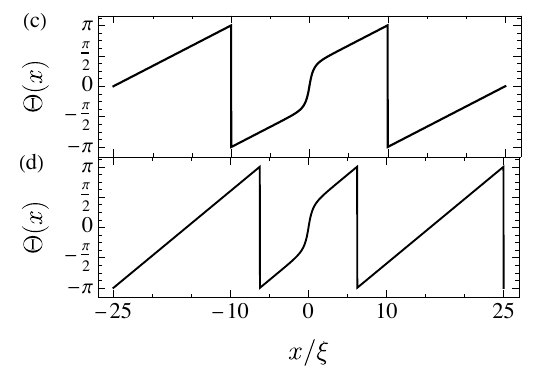}
	\caption{Phase of the condensate wave function for a) several values of the system momentum in the range $0<p<\pi \hbar n_0$, b) several values of the system momentum in the range $\pi \hbar n_0<p<2\pi \hbar n_0$, c) for $p=7\pi \hbar n_0/2$ and d)  $p=11\pi\hbar n_0/2$. For illustrative purposes, the system length is set to $L=50\xi$.} 
	\label{fig0}
\end{figure*}

The phase of the condensate wave function as a function of the spatial coordinate and the total system momentum  is shown in Fig.~\ref{fig0}. The total increase of the phase over the system length is $2k\pi$ for the momentum in the interval $\pi \hbar n_0 (2k-1) \leq p\leq \pi\hbar n_0(2 k+1)$. The physical mechanism behind this phenomenon is the following. Increasing the momentum of the system from zero to $\pi\hbar n_0$, the phase drop over the impurity increases from zero to its maximal value of $\pi$. At momentum $(\pi\hbar n_0)^{-}$, the impurity velocity is $0^+$ and the stationary state is a dark soliton. This phase drop over the impurity is compensated in the rest of the system, such that the total phase variation along the system is zero, see Fig.~\ref{fig0}(a). Note that $\theta$, given by \eq{eq:theta}, is a discontinuous function of the impurity velocity, and the abrupt change happens at $V=0$.
At $(\pi\hbar n_0)^+$, the impurity velocity is $0^-$, the phase drop over the impurity is $-\pi$ and the phase increase over the rest of the system is $\pi$. As a result, the total phase increase along the system is $2\pi$.
Increasing further $p$ from $\pi\hbar n_0$ to $2\pi \hbar n_0$, $\theta$ increases to $0$, while the total phase increase over the system remains $2\pi$, see Fig.~\ref{fig0}(b). Since $\theta$ is a periodic function of momentum, the process of growing of the phase variation over the system length repeats periodically, as illustrated in Fig.~\ref{fig0}.
Note that $\Theta$ shown in Fig.~\ref{fig0} is defined in the interval $ [-\pi, \pi)$, and thus the number of abrupt vertical drops of $2 \mathrm{sgn}(k)\pi$ is actually given by $|k|$.

\section{Bloch oscillations \label{sec:general}}

In this section we study the impurity dynamics in the presence of a force. We assume that the impurity and the bosons are initially in their zero-momentum ground state. Thus, the initial system momentum is $p=0$. Then, the boson-impurity coupling is turned on, and the impurity starts experiencing a constant force $F$. We numerically solve  \eq{eq:mean-field1}  with periodic boundary conditions. The discretization of time and space is performed using the protocol of conservative finite difference scheme, that is implemented fully implicitly. Then, the obtained system of nonlinear algebraic equations is solved iteratively \cite{GPE_discretization}. 
The first-order upwind scheme has been implemented in the discretization of the drift term, in order to respect the causality of the impurity propagation \cite{upwind_scheme}. 
The considered system size is sufficiently long, such that the density waves do not reach the boundaries of the system during the reported time evolution. 

In the absence of a force, we find that the dispersive density shock waves are emitted symmetrically on both sides of the impurity, taking away the background particles. A depletion in the boson density is created at the impurity position, while the impurity velocity remains zero. After the relaxation, the bosons in the vicinity of the impurity arrive at the ground state  (\ref{densityStationary}) for $V=0$. 

Next, we consider the system in the presence of an external force acting on the impurity. If the force is infinitesimal, the finite-momentum ground state considered in Sec.~\ref{sec:Stationary} at $p=F t$ is realised at each $t$. Thus, the $2\pi \hbar n_0$ periodicity of the ground-state energy and the impurity velocity as a function of momentum, insure that the impurity performs a periodic motion in space around a fixed position. These are the so-called Bloch oscillations. Their period is $T=2\pi \hbar n_0/F$. Since the maximal absolute value of the impurity velocity is given by the critical velocity, the latter determines the amplitude of the impurity velocity oscillations.
No energy is transferred to the system on average and no excitations are generated in the adiabatic approximation, $F\to 0^+$. In this case, the momentum given to the system over the time period is carried by the supercurrents, as explained in Sec.~\ref{sec:Stationary}.

What happens at a finite force? One would expect that the main obstacle for the realization of the Bloch oscillations in that case is the gapless continuum of excited states above the ground state. The elementary excitations consist of density waves and solitons. However, we find that the excitations are emitted periodically such that
the impurity momentum $M V(t)=F t-p_{\mathrm{exc.}}(t)-p_{\mathrm{p.grad.}}(t)-p_{\mathrm{hole}}(t)$ remains finite, and exhibits periodic oscillations with time period $T$ around a nonzero mean value. Here, all the contributions are considered in the laboratory frame. The momentum of the emitted excitations till a time $t$ is $p_{\mathrm{exc.}}(t)$, while $p_{\mathrm{p.grad.}}(t)$ denotes the contribution from the gradient of the phase at time $t$, not carried by the excitations nor the local hole. The momentum of the local hole around the impurity is $p_{\mathrm{hole}}$. The latter is also a periodic function in time with the period $T$. 
Interestingly, we find that the period of oscillation satisfies $T\approx 2\pi \hbar n_0/F$ in a wide range of forces. Later in this section, we will discuss the deviations from this expression as the force increases. Thus, the momentum transferred to the system over the time period is $ 2\pi\hbar n_0$. Since the impurity velocity and $p_{hole}$ are periodic function of time, the momentum of emitted excitations during the period $T$ satisfies $p_{\mathrm{exc.}}(T)+p_{\mathrm{p.grad.}}(T)\approx 2\pi\hbar n_0$. Furthermore, the total increase of the phase variation along the system over the period is $2\pi$. Thus, $p_{\mathrm{p.grad.}}(T)=\hbar n_0 (2\pi-\theta_{exc.})$, where $\theta_{exc.}$ is the total phase increase across the excitations. The total momentum of the solitons and the density waves is $\hbar n_0\theta_{exc.}$ at not too big forces.

The impurity velocity takes the form
\begin{align}
V(t) = V_0 + V_B f(t),
\label{Voft}
\end{align}
once the oscillations are established. Here $V_B$ is the amplitude of the oscillations, and $f(t)$ is a periodic function in time with the period $T$. We defined $V_0=(V_{\mathrm{max}}+V_{\mathrm{min}})/2$ and $V_B=(V_{\mathrm{max}}-V_{\mathrm{min}})/2$, where $V_{\mathrm{max}}$ and  $V_{\mathrm{min}}$ are the maximal and the minimal impurity velocity reached during the oscillations, respectively.
Note that $V_0$ and $V_B$ are time-independent. They depend on the force $F$ and other system parameters. 
The impurity velocity averaged over one period gives the drift velocity
\begin{align}\label{eq:drift}
	V_d=V_0+ \frac{V_B}{T} \int_{0}^{T} f(t) \mathrm{d}t.
\end{align}
In the following, we will refer to the impurity dynamics (\ref{Voft}) as the Bloch oscillations, although these are the  drifted Bloch oscillations. 
The energy pumped into the system per time period is 
\begin{align}\label{FVT}
\Delta E=F V_d T .
\end{align} 
It is actually the energy, evaluated in the laboratory reference frame, of emitted solitons, dispersive density shock waves and density waves, as well as of the phase variations produced over one period and not included in the excitations. At small $F$, the drift velocity is a linear function of the force,  $V_d = \sigma F$, where $\sigma$ denotes the impurity mobility. The latter can be expressed using only the properties of the finite-momentum equilibrium ground state \cite{AnnalsKamenev}. 
The characteristics of the Bloch oscillations as a function of the system parameters will be discussed below, as well as in the following sections.

\begin{figure*}[t!]
	\centering
	\includegraphics[width=0.66\columnwidth]{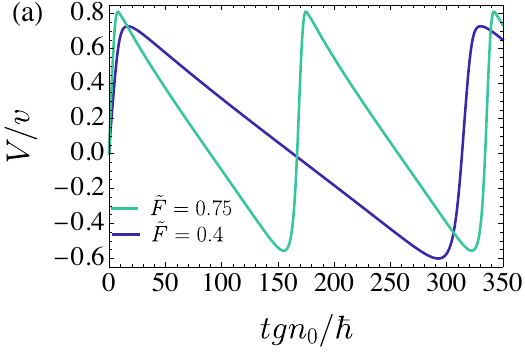}
	\includegraphics[width=0.64\columnwidth]{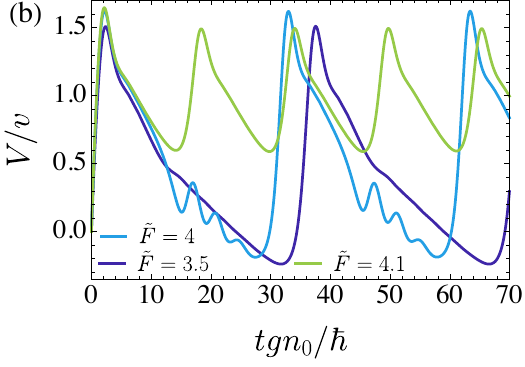}
	\includegraphics[width=0.60\columnwidth]{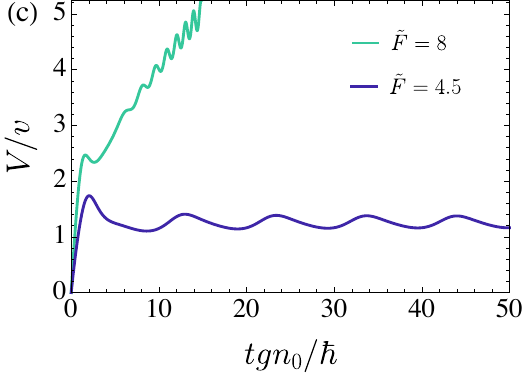}\label{fig:2b}
	\includegraphics[width=0.64\columnwidth]{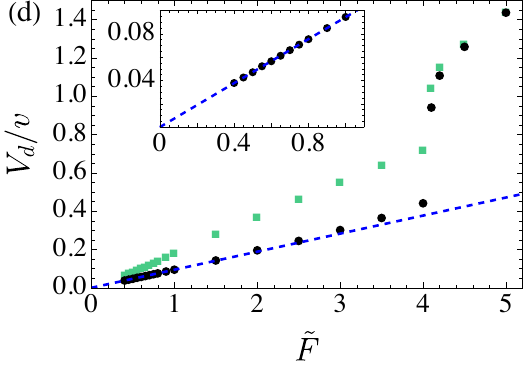}
	\includegraphics[width=0.64\columnwidth]{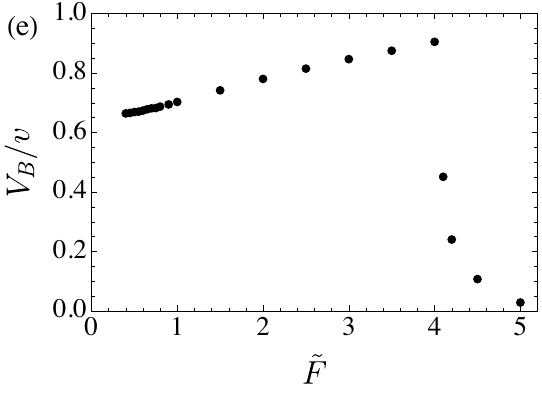}
	\includegraphics[width=0.64\columnwidth]{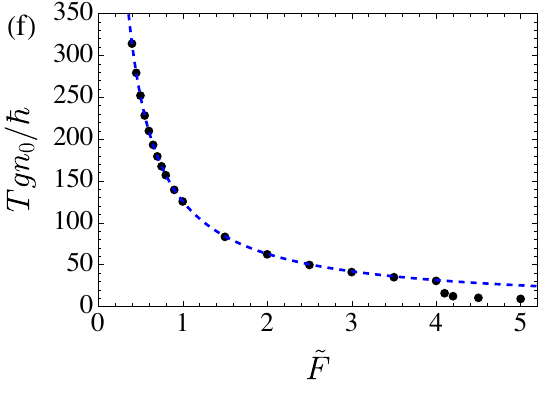}
	\caption{Time evolution of the impurity velocity for $M=3m$, $\tilde{G}=0.5$ and $\gamma = 1/400$ in the a) small, b) intermediate and c) large-force regime. d) The dimensionless drift velocity $V_d/v$ is shown by the black points and $V_0/v$ by the green squares as a function of the dimensionless force $\tilde{F}$. The linear fit of the drift velocity is represented by the dashed line, $V_d/v = \tilde{\sigma} \tilde{F}$, and gives us the dimensionless mobility $\tilde{\sigma}= 0.094$. e) The dimensionless Bloch oscillation amplitude $V_B/v$ and f) the dimensionless time period $gn_0 T/\hbar$ are shown as a function of the dimensionless force $\tilde{F}$. The dimensionless time period is fitted with the prediction $gn_0 T/\hbar = 2 \pi/(\tilde{F} \sqrt{\gamma})$, shown by the dashed line.}
	\label{fig:general}
\end{figure*}

\begin{figure*}[t]
	\centering
	\includegraphics[width=2\columnwidth]{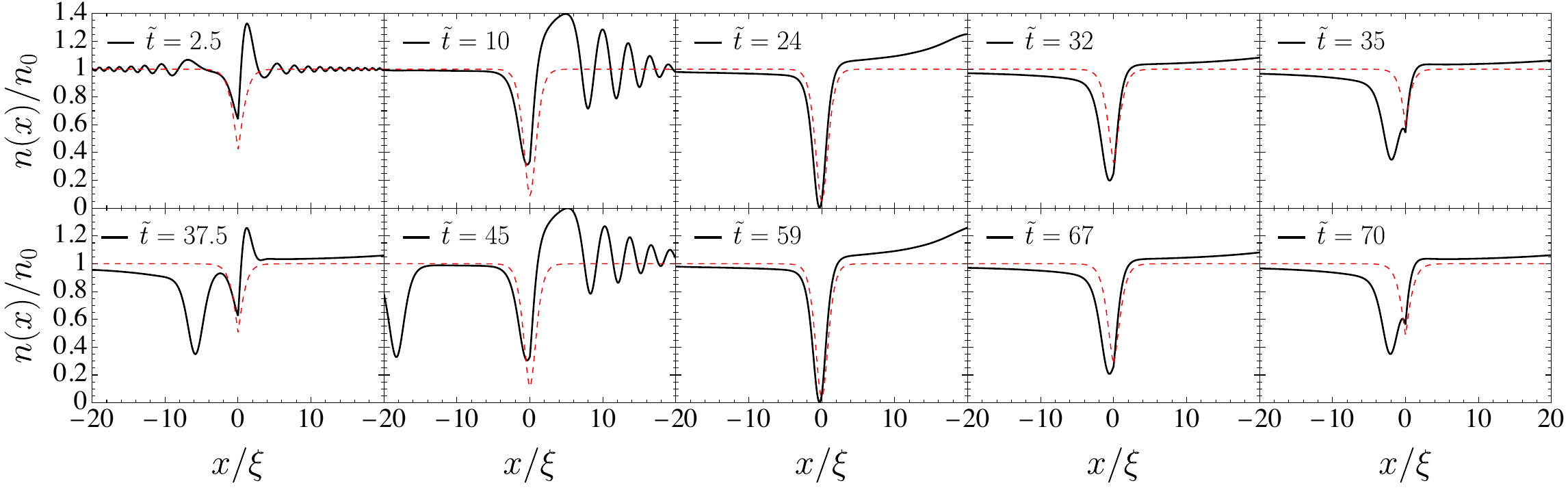}
	\caption{Time evolution of the boson density in the reference frame defined in Sec.~\ref{sec:model}, for $\tilde{G}=0.5,M=3m,\tilde{F}=3.5$ and $\gamma=1/400$.  The dimensionless time is defines as $\tilde{t}=t g n_0/\hbar$. The ground-state configuration, \eq{densityStationary}, for the momentum $p=F t$ is shown by the dashed line. The impurity velocity reaches maximal value at approximately $2.4$ and $37.6$, its minimal value at $31.7$ and $66.6$, and zero at $24.1$, $34.4$, $58.9$, and $69.3$.}
	\label{fig:density}
	\end{figure*}

Having explained the general physical picture, we now present supporting results. Fig.~\ref{fig:general} shows the Bloch oscillations of the impurity velocity in time, as well as the drift velocity, the Bloch amplitude, and the time period as a function of the driving force for the case of $\tilde{G}=0.5$, $M=3m$ and $\gamma=1/400$. We define the dimensionless impurity strength as $\tilde{G}=G/\hbar v$, and the dimensionless force as $\tilde{F}=F \xi/g n_0$. 
By increasing the driving force $F$, we identify different regimes of the impurity dynamics. 
In order to understand the underlying mechanism, we first analyze the time evolution of the boson density in the reference frame defined in Sec.~\ref{sec:model}, for a fixed dimensionless force, see Fig.~\ref{fig:density} for $\tilde{F}=3.5$. As the impurity-bath coupling is switched on at $t=0$, the density dispersive shock waves start forming. They move away from the impurity, taking away some particles. As a result, a depletion in the boson density is created around the impurity position. The impurity reaches its maximal velocity at the time $\tilde{t}=t g n_0/\hbar=2.4$. As time progresses, the impurity velocity slowly decreases and reaches zero at $\tilde{t}=24.1$, while simultaneously the depth of the local hole increases and reaches  a complete depletion.
During this process, density shock waves are continuously emitted. The density profile differs from the ground-state density profile (\ref{densityStationary}) evaluated at the momentum $p=F t$, shown by the dashed line in Fig.~\ref{fig:density}. At $\tilde{t}=31.7$ the impurity velocity reaches its minimal value, and then undergoes a very fast deceleration, see Fig.~\ref{fig:general}b. While the absolute value of the impurity velocity decreases to zero, the depletion at the impurity position rapidly decreases, leading to the formation of a soliton, see Fig.~\ref{fig:density}. The soliton moves away, carrying a part of the depletion from the local hole. At $\tilde{t}=37.6$ the impurity velocity reaches its maximal value. Afterwards, further increase of the local boson depletion is accompanied by the emission of new density shock waves in front of the impurity. The described process is repeated periodically. As a result, one soliton is emitted per period. 

The periodicity of the impurity velocity is established after it reaches its first zero value, at $\tilde{t}=24.1$. The reason behind this is the first soliton emission that takes place while the impurity velocity increases from its first minimal value to zero. This process causes a small decrease of the time needed to go from the second $V_{\mathrm{max}}$ to the second $V_{\mathrm{min}}$ with respect to the first ones. The time period of oscillations at $\tilde{F}=3.5$ deviates by $3\%$ from the value $2\pi \hbar n_0/F$.

\begin{figure*}
	\includegraphics[width=0.65\columnwidth]{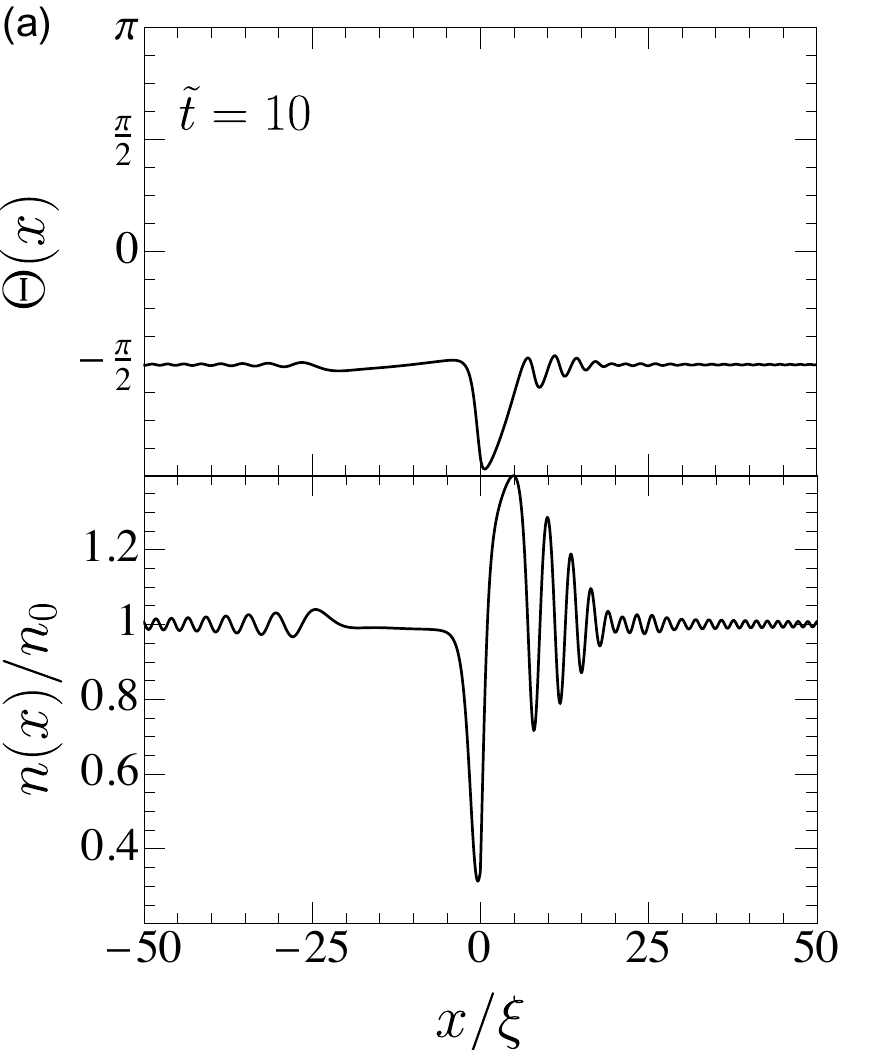}
	\includegraphics[width=0.65\columnwidth]{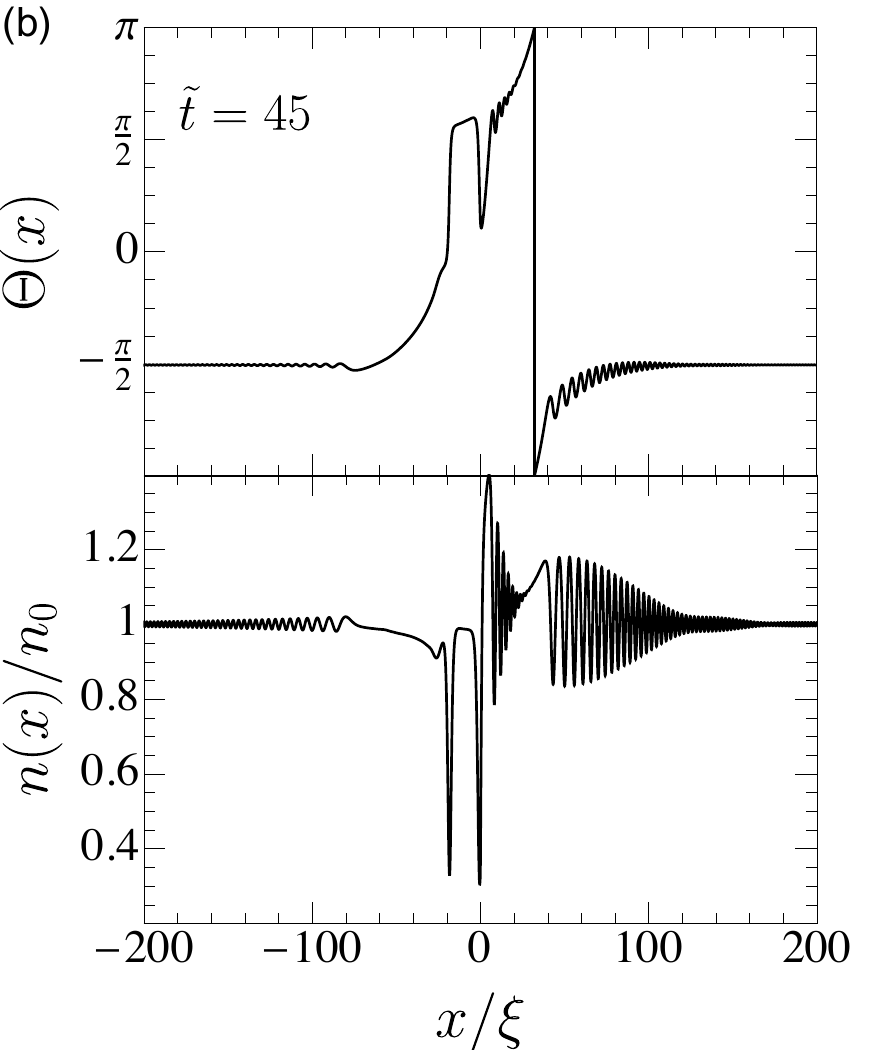}
	\includegraphics[width=0.65\columnwidth]{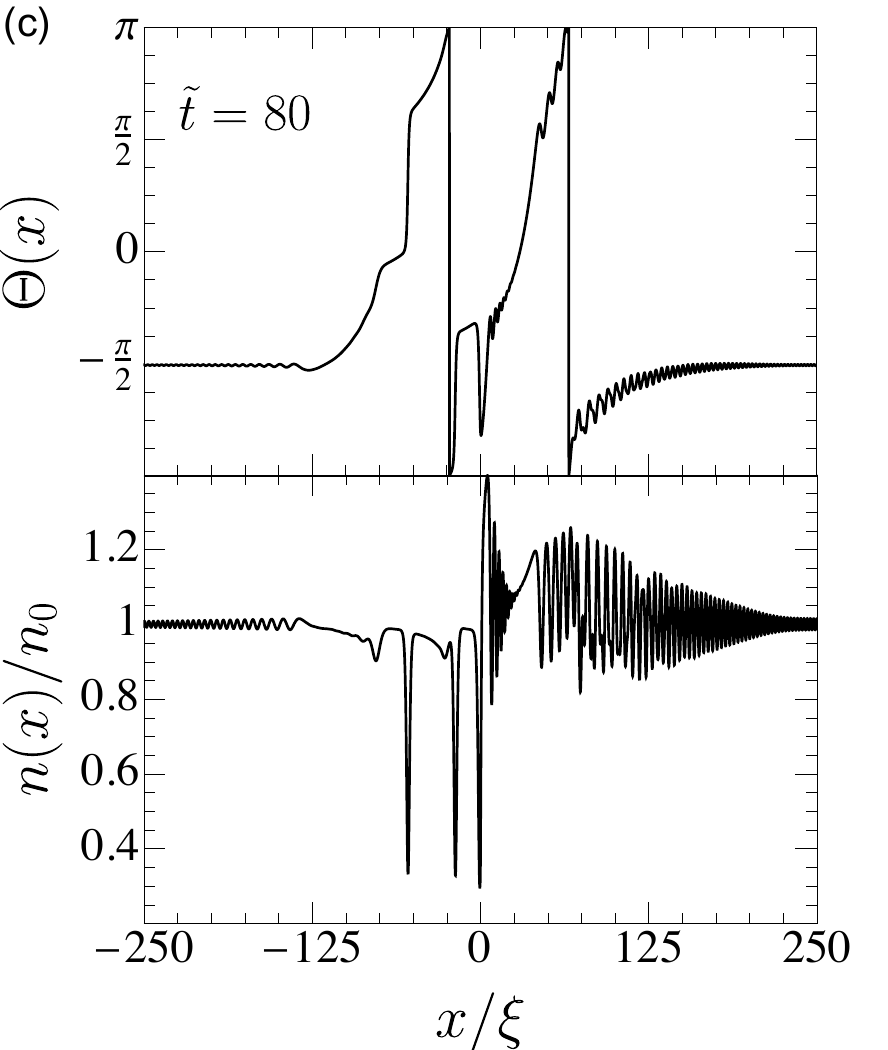}
	\caption{Phase (top row) and density profile (bottom row) of the bosons in the reference frame defined in Sec.~\ref{sec:model} for $\tilde{G}=0.5,M=3m,\tilde{F}=3.5$, and $\gamma=1/400$ at different times: (a) $\tilde{t}=10$, (b)  $\tilde{t}=45$, and (c)  $\tilde{t}=80$. The time period of the impurity velocity oscillations is $\tilde{T}=34.8$.}
	\label{fig3}
\end{figure*}

Next, we study the time evolution of the phase of the condensate wave function. The top and the bottom rows of Fig.~\ref{fig3} show the phase and the boson density, respectively, in the reference frame defined in Sec.~\ref{sec:model} at a given time. 
As time passes, the firstly emitted density waves move away from the impurity and the region between them, where most of the phase variations take place, increases in time. Additionally, small oscillations of the phase occur along the firstly emitted density waves. We note that the absolute value of the phase gradient is bigger around the impurity and the solitons 
than in the rest of the system. At later times, the boson density profile shows the interference of the density waves emitted at different times, see Fig.~\ref{fig3}(c).
The phase difference over the system length increases for $2\pi$ over one time period. As a result, an additional vertical line along which the phase drops for $2\pi$ appears per period, see Fig.~\ref{fig3}. 
 
At a finite driving force, the above described physical picture replaces the one shown in Fig.~\ref{fig0}, valid in the case of an infinitesimal force, where the constant phase gradient takes place along the \emph{whole} system. However, in the limit of vanishingly small force, the time period becomes infinitely big, and the firstly emitted density waves are then infinitely far away from each other and we recover the $F\to 0^+$ behaviour. 
Decreasing the driving force, the emitted density waves and solitons become less energetic. At $F\to 0^+$, the boson density around the impurity approaches the ground-state density (\ref{densityStationary}) at $p=F t$. The impurity velocity as a function of the total system momentum $p=F t$ tends to the velocity of the ground state as the force decreases, as shown in Fig.~\ref{fig1}. However, note that increasing the force, the relative error of the time period with respect to $2\pi \hbar n_0/F$ increases. At $\tilde{F}=2$ the period is $1\%$ shorter than the expected one. Moreover, Fig.~\ref{fig1} shows that the shape of oscillations changes with the force. This is even more pronounced at bigger forces displayed in Fig.~\ref{fig:general}.

The drift velocity, the Bloch amplitude, and the time period as a function of the force are shown in the bottom row of Fig.~\ref{fig:general}. The drift velocity is an increasing function of the driving force. At small $F$, the drift velocity is a linear function of the force,  $V_d = \sigma F$. We define the dimensionless mobility as $\tilde{\sigma} = \sigma mgn_0/\hbar$, and for the parameters of Fig.~\ref{fig:general}, we evaluate it to be 
$\tilde{\sigma} = 0.094$.  For larger forces, $V_d$ is bigger than $\sigma F$, i.e., it exhibits a super-linear behavior. Note that both terms in the drift velocity (\ref{eq:drift}) are relevant. The first contribution is shown by the green squares in Fig.~\ref{fig:general} (d). The second contribution is negative. Even in the limit of small forces, the second term in \eq{eq:drift} matters and gives an important contribution to the impurity mobility. For $\tilde{F}=4$, the oscillations of the impurity velocity display small ripples, which, at $\tilde{F}=4.1$, transform into a new type of oscillations, giving rise to an abrupt increase of the drift velocity. However, the Bloch oscillations still persist. We discuss this regime in more detail below. 

The amplitude of the impurity velocity oscillations, $V_B$, is a nonmonotonic function of the driving force. It tends to the critical velocity in the limit of an infinitesimal force. The amplitude $V_B$ first increases with $F$ and reaches it maximal value. Then it abruptly drops, and becomes a decreasing function of $F$. It tends to zero at bigger forces, in the regime where the Bloch oscillations persist.

Increasing the driving force, the oscillations of the impurity velocity become faster. The time period decreases as $T = 2\pi \hbar n_0/F$. The latter expression holds even at bigger forces.  However, the relative error with respect to this prediction is around $0.8\%$ at $\tilde{F}=1$ and grows with $\tilde{F}$ up to $3\%$ at $\tilde{F}=4$. 
At $\tilde{F}=4.1$, the period abruptly decreases and becomes twice smaller than the expected one.  At $\tilde{F}\geq 4.1$, the period is not anymore given by $2\pi \hbar n_0/F$, signaling the breakdown of the established mechanism. The periodic emission of solitons still takes place in this regime and leads to 
stable impurity velocity oscillations. However, the total phase drop over the  system length remains zero in time. Fig.~\ref{fig:general}(c) shows that the Bloch oscillations disappear at $\tilde{F}=8$ and a new kind of the dynamics arises, where the impurity velocity keeps increasing in time. The superimposed oscillations visible in Fig.~\ref{fig:general}(c) at $\tilde{F}=8$ disappear at longer times. 

\begin{figure*}
	\centering
	\includegraphics[width=\linewidth]{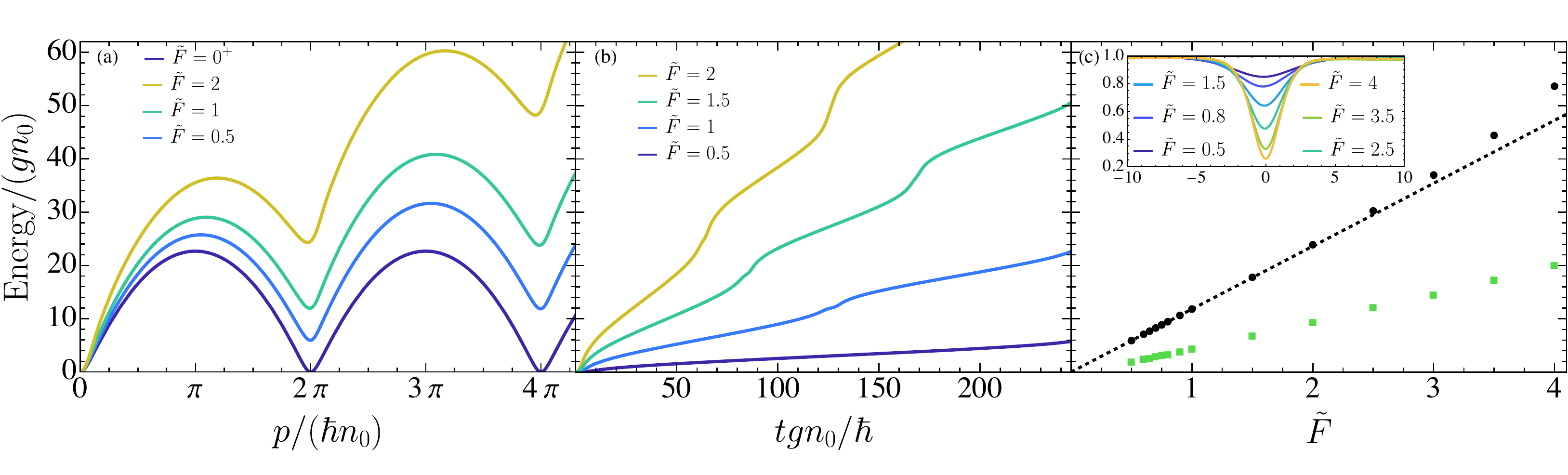}
	\caption{(a) Energy of the system $E(p)$ measured with respect to the energy of the state at $p=0$, as a function of the momentum $p=F t$ for three different values of $\tilde{F}=0.5, 1$ and $2$. Here $M=3m$, $\tilde{G}=0.5$, and $\gamma=1/400$. The lowest-energy  curve shows the energy of the ground state of the system at finite momentum $p$, measured with respect to the energy of the state at $p=0$. This is the energy of the system in the adiabatic approximation $E_{\mathrm{ad}}(p)$, i.e., obtained for an infinitesimal force. (b) The energy of emitted excitations, $E(F t)-E_{\mathrm{ad}}(Ft)$, as a function of time for $\tilde{F}=0.5, 1, 1.5$ and $2$. (c) The energy of all emitted excitations and the energy of solitons only, emitted per period, as a function of the driving force are shown by the black dots and the green squares, respectively. The inset shows the density profile of emitted solitons at different forces, measured in units of $n_0$ as a function of the spatial coordinate expressed in units of $\xi$. The dashed line is the linear fit of the total energy of the excitations per period in the low-$\tilde{F}$ regime. It reads as $\sigma F^2 T/(g n_0) =2\pi \tilde{F}\tilde{\sigma}/\sqrt{\gamma}=11.8 \tilde{F}$. }
	\label{fig:energy}
\end{figure*}

Figure \ref{fig:energy} (a) shows the total energy of the system, measured with respect to the energy of the state realized at $p=0$, as a function of the total momentum $p=F t$. It indicates how far the system is from the ground-state configuration in the presence of a finite force. The ground-state energy is the lowest-energy curve shown in Fig.~\ref{fig:energy} (a). The energy of the emitted excitations as a function of time is presented in Fig.~\ref{fig:energy} (b) for several values of the force. In Fig.~\ref{fig:energy} (c), the black circles show the energy (\ref{FVT}) pumped into the system over one period, in a wide region of forces, going beyond the linear $V_d$ dependence on $F$. The dashed line in Fig.~\ref{fig:energy} (c) shows the contribution $2\pi \tilde{F}\tilde{\sigma}/\sqrt{\gamma}$ that corresponds to the small-$F$ behaviour of $F V_d T/g n_0$. 
The emitted solitons become more energetic as the force increases, their energy is shown by the green rectangles in Fig.~\ref{fig:energy} (c). The inset shows the evolution of the soliton density profile as a function of $F$. Contrary to the expectations \cite{AnnalsKamenev}, solitons are emitted even though the drift impurity velocity is smaller, even considerably smaller, than the sound velocity. They are actually present at any $F$, and their maximal depletion increases continuously with $F$.
The maximal force, $F_{\textrm{max}}$, under which action the Bloch mechanism exists was estimated to be given by the ratio of the sound velocity and the mobility \cite{AnnalsKamenev}. For the parameters of Fig.~\ref{fig:strongGVelocity}, the predicted maximal force is $\tilde{F}_{\text{max}} = \sqrt{1+m/M}/\tilde{\sigma} \approx 12$. This prediction overestimates numerically calculated $\tilde{F}_\textrm{max}$ by around $50\%$ here, but it underestimates significantly $\tilde{F}_\textrm{max}$ at strong coupling case studied in Sec.~\ref{sec:strong}.

\section{Strongly-coupled impurity\label{sec:strong}}

\begin{figure}
	\centering
	\includegraphics[width=\columnwidth]{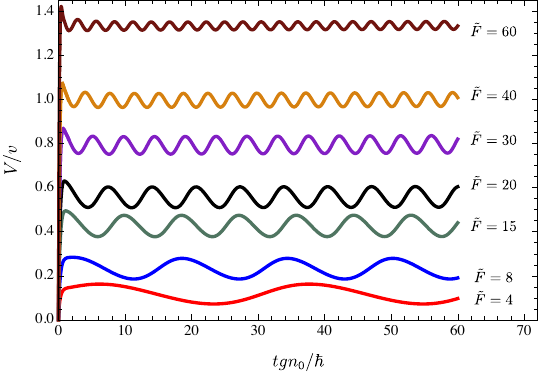}\\
		\caption{Impurity velocity as a function of time for $\tilde{G}=15$ and $M=3m$ for different values of driving force. Here $\gamma = 1/400$.}
	\label{fig:strongGVelocity}
\end{figure}

\begin{figure*}
	\centering
	\includegraphics[width=0.65\columnwidth]{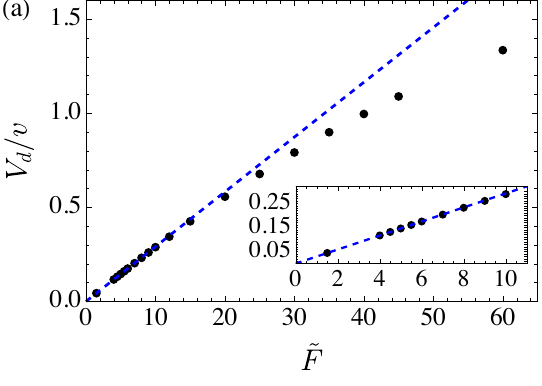}\phantom{aa}
	\includegraphics[width=0.669\columnwidth]{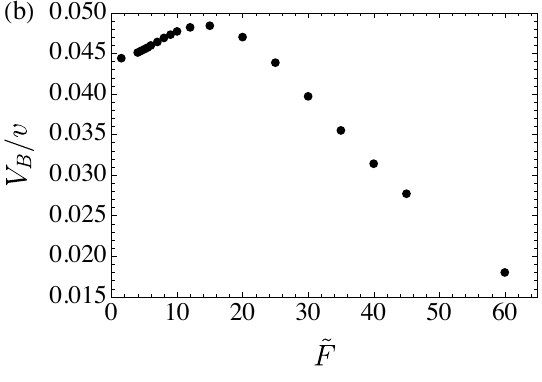}\phantom{aa}
	\includegraphics[width=0.64\columnwidth]{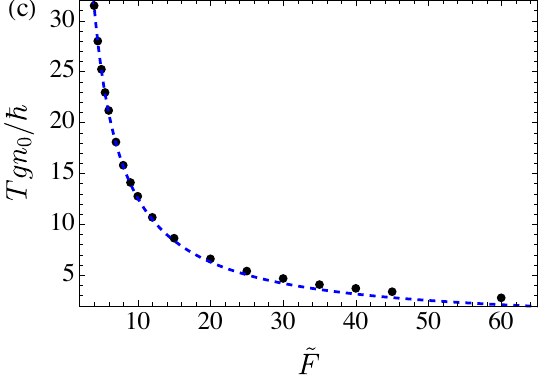}
	\caption{ (a) Drift velocity, (b) Bloch oscillation amplitude and (c) time period of the oscillations as a function of the dimensionless force $\tilde{F}$. Here $\tilde{G}=15$, $M=3m$, and $\gamma = 1/400$. The linear fit in the drift velocity (dashed line) $V_d/v = \tilde{\sigma} \tilde{F}$ gives us the dimensionless mobility $\tilde{\sigma} = 0.029$. The dimensionless time period is fitted with the prediction $gn_0 T/\hbar = (2 \pi)/(\tilde{F} \sqrt{\gamma})$ (dashed line). }
	\label{fig:strongG}
\end{figure*}

In this section, we investigate a strongly-coupled impurity, $\tilde{G} >> 1$, that exhibits a different behavior than the general case studied in Sec.~\ref{sec:general}. We show the impurity velocity time evolution for different values of the driving force for $\tilde{G}=15$, $M=3m$, and $\gamma=1/400$ in Fig.~\ref{fig:strongGVelocity}. Note that here the Bloch oscillations have a more regular sinusoidal shape. As a result, the second term in \eq{eq:drift} is negligible. The Bloch amplitude $V_B$ is strongly suppressed. Just like in a general case, increasing the driving force the Bloch amplitude and the drift velocity increase, whereas the time period decreases, as shown in Fig.~\ref{fig:strongG}. The mobility in the strong coupling case is also considerably reduced, which we evaluate to be $\tilde{\sigma} = 0.029$. Moreover, the drift velocity $V_d$ increases sub-linearly at stronger forces, contrary to the case of a moderate $\tilde{G}$ shown in Fig.~\ref{fig:general}. In the strong coupling case, the Bloch amplitude grows slowly with $F$, reaches its maximal value and then it smoothly crosses over to a monotonically decreasing regime. There is no abrupt change of the drift velocity, as observed in the case of a moderate impurity coupling.

\begin{figure}
	\centering
	\includegraphics[width=0.49\linewidth]{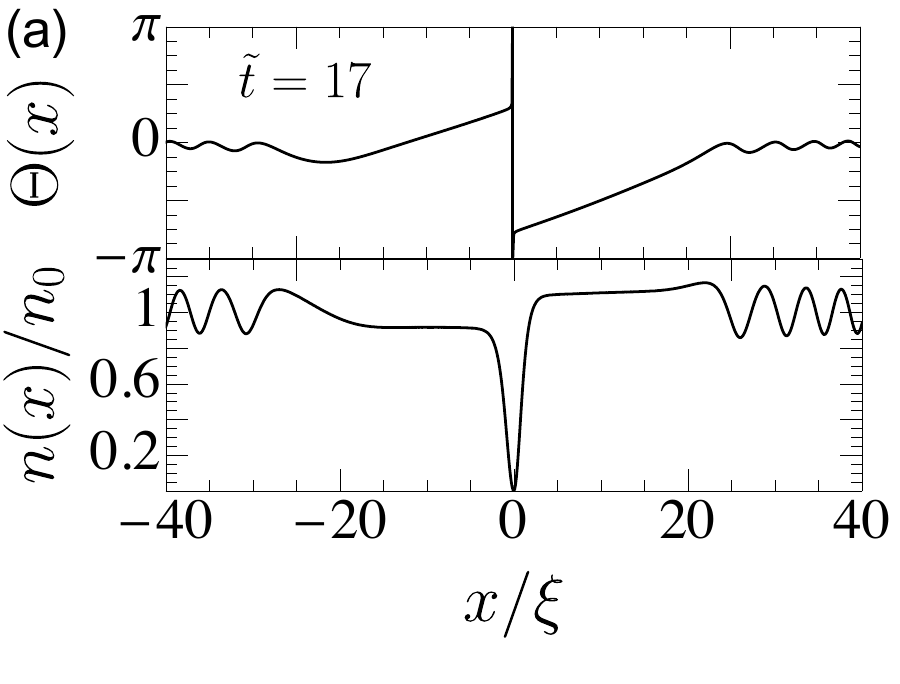}
	\includegraphics[width=0.49\linewidth]{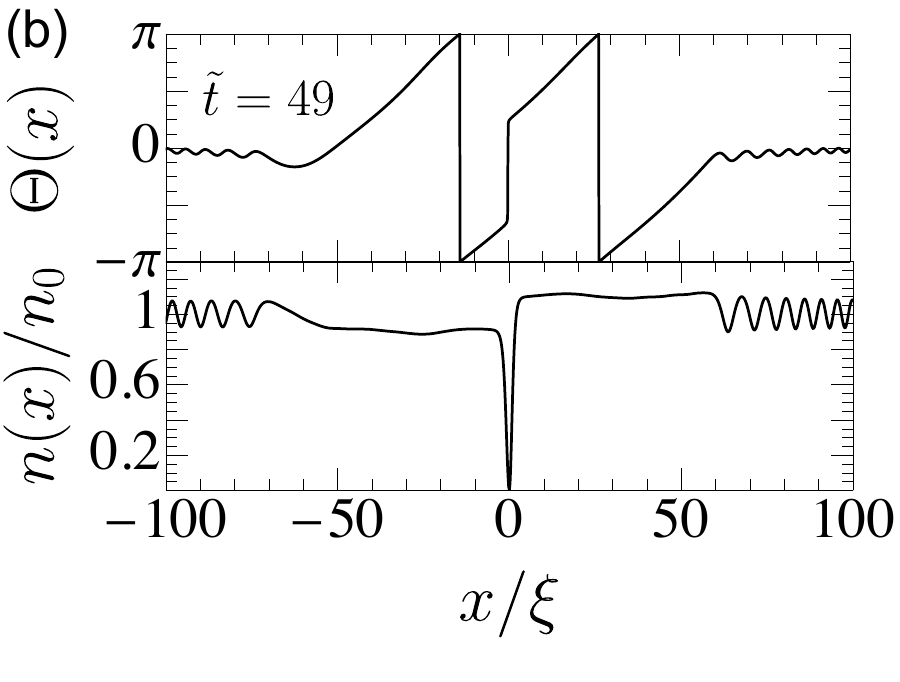}
	\caption{Phase and density profile of the bosons in the reference frame defined in Sec.~\ref{sec:model} for $\tilde{G}=15, M=3m, \tilde{F}=4$ taken at (a) $\tilde{t}=17$ and at (b) $\tilde{t}=48$. Here $\gamma=1/400$ and $T=31.5$. }
	\label{fig:densityStrong}
\end{figure}

The time evolution of the boson density in the reference frame defined in Sec.~\ref{sec:model} is shown in Fig.~\ref{fig:densityStrong}. After emitting dispersive density shockwaves, a local complete depletion is formed at the impurity position, and remains present throughout the process of velocity oscillations. As a result, the emission of solitons and additional shock waves does not take place here. Furthermore, the boson density is lower and higher than the mean boson density behind and in front of the impurity, respectively, in the region between the impurity and the firstly emitted density waves. This region grows in time as the density waves move away from the impurity. Thus, the energy pumped into the system over one period goes mostly into further expansion of this region, i.e., into the cost of the boson-boson interaction energy in this region.
Note that now the phase varies in space in a more regular manner, while its periodic increase of $2\pi$ over the system length also occurs here, see Fig.~\ref{fig:densityStrong}. The time period as a function of the driving force shows a deviation from the expected result $2\pi\hbar n_0/F$ (showed by the dashed line in Fig.~\ref{fig:strongG}) at bigger forces. Contrary to a weak and moderate $\tilde{G}$, here the period is bigger than the expected one. This deviation grows with $\tilde{F}$. It is $3\%$ for $\tilde{F}=15$, $11.5\%$ for $\tilde{F}=30$, and $32\%$ for $\tilde{F}=60$ of the numerically obtained value.

The case of a strong coupling regime was studied in Ref.~\cite{AnnalsKamenev}. 
The authors provide an analytical result for the mobility of the impurity, which in our dimensionless units is given by $\tilde{\sigma} = \sqrt{\gamma(1+m/M)}/2$. For the aforementioned system parameters, it takes the value $\tilde{\sigma} = 0.029$. Our results show an excellent agreement with this prediction. 
It is also claimed that the Bloch oscillation mechanism gets ruined as the impurity drift velocity reaches the sound velocity and the Cherenkov radiation gets triggered. Thus a maximal force under which action the Bloch mechanism is present was estimated to be given by the ratio of the sound velocity and the mobility \cite{AnnalsKamenev}. For the parameters of Fig.~\ref{fig:strongGVelocity}, the predicted maximal force is $\tilde{F}_{\text{max}} = \sqrt{1+m/M}/\tilde{\sigma} \approx 40$. However, we find that the Bloch mechanism exists at much larger dimensionless forces, even of the order of several hundreds. At bigger $F$, the amplitude of the velocity oscillations becomes zero, while the drift velocity remains constant in time. Moreover, the soliton emission remains absent and the Bloch oscillations exist even though the drift velocity overcomes the critical velocity, as well as the mass-scaled sound velocity $v\sqrt{1+m/M} = 1.155 v$ of the bosonic medium. Furthermore, at $F_{\text{max}}$, the system is far from the linear $V_d=\sigma F$ regime.

Using the Josephson junction description of the impurity, the drift velocity and the oscillation time period are evaluated analytically up to the second order in $F/F_{\text{max}}$ in Ref.~\cite{AnnalsKamenev}. We find that these predictions neither describe the deviation of the drift velocity from the linear $\sigma F$ law nor the time period from $2\pi\hbar n_0/F$ dependence. These predictions do not show a better matching with our numerical results even if $F_{\text{max}}$ is used as a fitting parameter.

\section{Dynamics as a function of the impurity-boson coupling, the impurity mass and the boson-boson interaction \label{Sec:dependence}}

\begin{figure}
	\centering
		\includegraphics[width=\columnwidth]{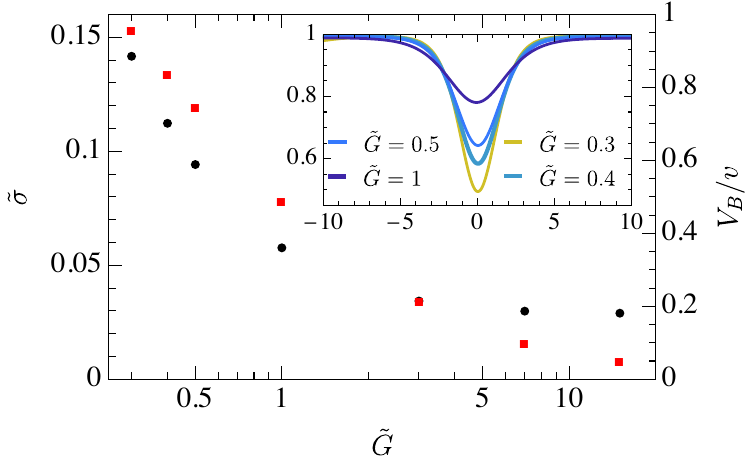}
\caption{Dimensionless mobility, $\tilde{\sigma} = \sigma mgn_0/\hbar$, as a function of the dimensionless impurity-boson coupling for $M=3m$ and $\gamma=1/400$ is shown by the black dots. For the same parameters, the Bloch amplitude  as a function of $\tilde{G}$ is shown by the red squares for $\tilde{F}=1.5$. The inset shows the density profile of the emitted solitons at different $\tilde{G}$ for $\tilde{F}=1.5$, measured in units of $n_0$ as a function of the spatial coordinate expressed in units of $\xi$.}
	\label{fig:G}
\end{figure}

In this section, we study the properties of the Bloch oscillations as a function of $\tilde{G}$, $M/m$ and $\gamma$. To ensure that the mobility has been properly extracted, we use the ratio of the critical velocity $v_c$ and the Bloch amplitude corresponding to the smallest driving force $V_{B,\mathrm{min}}$ as a control parameter (recall that for $F \to 0^+$, $V_B \to v_c$). For all the results shown here, we have $0.1 \% \leq (V_{B,\mathrm{min}} - v_c)/v_c \leq 1.8 \%$.  Furthermore, as shown in previous sections,  the regime of the linear dependence of the drift velocity on $F$ holds in a wide interval of forces.

Having already discussed in the two previous sections the cases of a moderately and a strongly coupled impurity, we first focus on $\tilde{G}$ dependence. As the impurity-bath coupling increases, the motion of the impurity becomes more restricted and thus its mobility decreases, as shown in Fig.~\ref{fig:G}. As $\tilde{G}$ is increased from $0$ to $1$, $\sigma$ drops rapidly, and at bigger $\tilde{G}$, it shows a slower decline.
The mobility saturates to $\sqrt{\gamma(1+m/M)}/2$ at strong $\tilde{G}$. A similar behaviour was reported in Ref.~\cite{AnnalsKamenev}.

Note that the interval of driving forces where the drift velocity shows a linear dependence on $F$ shrinks down as $\tilde{G}$ decreases. We show a typical behaviour of the Bloch amplitude as a function of $\tilde{G}$, for a fixed force $\tilde{F}=1.5$, such that for all the shown values  of $\tilde{G}$, the drift velocity is well approximated by $\sigma F$ dependence. Both $V_d$ and $V_B$ are monotonically decreasing functions of $\tilde{G}$. In contrast to $V_d$, $V_B$ does not exhibit a slowly-varying regime at bigger $\tilde{G}$, and tends to zero as shown in Fig.~\ref{fig:G}. We study the response of the background bosons in the form of emitted solitons as a function of $\tilde{G}$ in the inset of Fig.~\ref{fig:G}. To capture their true shape, we show the solitons at a sufficiently long time when they are separated from the impurity and the other emitted excitations, although some interference effects remain visible in the tails of the solitons, especially at big $\tilde{G}$. The soliton depth, and thus also its energy, decreases with $\tilde{G}$, and at very strong coupling they tend to zero -- as discussed in Sec.~\ref{sec:strong}. 
The reason for this behaviour is a smaller variation of the boson depletion situated at the impurity position during the oscillations as $\tilde{G}$ increases. 
Thus, the emitted solitons take away smaller number of missing particle, becoming more shallow. For very strong $\tilde{G}$, the boson depletion at the impurity position remains total during the oscillations, leading to the absence of soliton emission.

\begin{figure}
	\centering
	\includegraphics[width=\columnwidth]{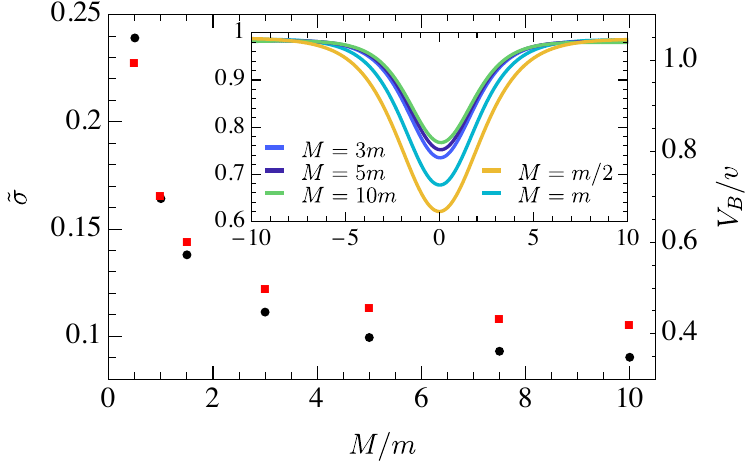}
	\caption{Dimensionless mobility, $\tilde{\sigma} = \sigma mgn_0/\hbar$, as a function of $M/m$ for $\tilde{G}=1$ and $\gamma=1/100$ is shown by the black dots. For the same parameters and $\tilde{F}=1$, the dimensionless Bloch amplitude $V_B/v$ is shown by the red squares. The inset shows the density profile of the emitted solitons at different ${M/m}$ for $\tilde{F}=1$, measured in units of $n_0$ as a function of the spatial coordinate expressed in units of $\xi$.}
	\label{fig:M}
\end{figure}

Next, we investigate the influence of the impurity-boson mass ratio $M/m$ on the dynamics. Figure \ref{fig:M} shows the dependence of the impurity mobility for $\tilde{G}=1$ and $\gamma=1/100$. The lighter the impurity is, the higher its mobility is. Note that the interval of driving forces where the drift velocity shows a linear dependence on $F$ shrinks down as $M/m$ decreases.  Figure \ref{fig:M} shows a typical behaviour of the Bloch amplitude as a function of $M/m$ for a fixed force $\tilde{F}=1$. For all the shown values  of $M/m$, the drift velocity is well approximated by $\sigma F$ dependence. 
Both $V_d$ and $V_B$ decrease as the impurity becomes heavier while the other parameters are kept fixed. 
The density profile of the periodically emitted solitons as a function of $M/m$ is shown in the inset of Fig.~\ref{fig:M}. Some interference effects between other solitons and shock waves are visible in the soliton tails, especially at big $M/m$. The soliton depth and its energy decrease with the impurity mass, as long as its mass remains smaller than the critical one. \footnote{For the definition of the critical mass see Refs.~\cite{PhysRevLett.108.207001} and \cite{Quench}.}

\begin{figure}
	\centering		
	\includegraphics[width=\columnwidth]{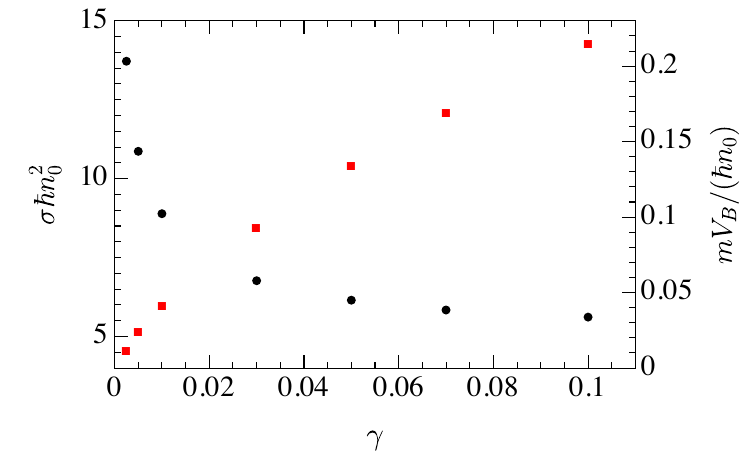}
\caption{Mobility as a function of the dimensionless boson-boson interaction strength $\gamma$ for $Gm/(\hbar^2 n_0)=0.15$ and $M=3m$ is shown by the black dots. For the same parameters, the Bloch amplitude $mV_B/\hbar n_0$ as a function of $\gamma$ is shown by the red squares for $(Fm)/(\hbar^2 n_0^3)=0.0025$.}
	\label{fig:gamma}
\end{figure}

In Fig.~\ref{fig:gamma}, we investigate the dependence of the Bloch oscillations on the boson-boson interaction strength. We fix $Gm/(\hbar^2 n_0)=0.15$ and $M=3m$, and vary $\gamma$.  Thus, $\tilde{G}$ decreases from $3$ at $\gamma=1/400$ to $0.47$ at $\gamma=1/10$. We now display $\gamma$-independent dimensionless quantities, such as  $\tilde{\sigma}/\gamma = \sigma \hbar n_0^2$ and $V_{B} m/\hbar n_0$. 
The impurity mobility decreases very rapidly at smallest values of $\gamma$ and exhibits a very slow decay at bigger values of $\gamma$, showing a tendency of saturation to a finite value. As $\gamma$ increases, $\tilde{G}$ decreases, and the saturation regime is in the weak-$\tilde{G}$ domaine. Thus, this tendency is in agreement with the prediction that at $\tilde{G}\ll 1$,  the dimensionful mobility is independent on the boson-boson interaction strength \cite{AnnalsKamenev}. 
Next, we study the behaviour of $V_B$. In Fig.~\ref{fig:gamma}, the force is fixed to be $(Fm)/(\hbar^2 n_0^3)=0.0025$. This value of the force is in the linear regime of $V_d$ dependence on $F$ for all gamma apart from the three smallest one. Thus, $V_d$ shows almost the same behaviour as $\sigma F$, it is a decreasing function of $\gamma$. On the other hand, the Bloch amplitude exhibits the opposite behaviour, it is an increasing function of $\gamma$. 

The impurity mobility was expressed using the properties of the finite-momentum equilibrium ground state in Ref.~\cite{AnnalsKamenev}. After adapting their result for the equation of motion (\ref{eq:mean-field1}), the dimensionless mobility, $\tilde{\sigma} = \sigma mgn_0/\hbar$, reads as
\begin{align} \label{eq:sigma}
\tilde{\sigma}={}&\frac{\sqrt{\gamma(1+m/M)}}{2\pi}\int_{-\pi \kappa}^{\pi \kappa}\frac{\mathrm{d} \tilde{p}}{1-\tilde{V}^2}\Bigg[\frac{\kappa}{2}\left(\frac{\partial\theta}{\partial\tilde{p}}\right)^2 \notag\\ &+\tilde{V}\left(\frac{\partial N}{\partial\tilde{p}}\right) \left(\frac{\partial\theta}{\partial\tilde{p}}\right) +\frac{1}{2\kappa} \left(\frac{\partial N}{\partial\tilde{p}}\right)^2\Bigg].
\end{align}
Here, we have introduced $\kappa=\sqrt{(1+m/M)/\gamma}$, $\tilde{V}=V/v\sqrt{1+m/M}$, and the dimensionless momentum $\tilde{p}=p\sqrt{1+m/M}/mv$. The phase drop $\theta$ across the impurity is given by \eq{eq:theta}, while the number of bosons $N$ expelled by the impurity is determined by \eq{eq:Nexpelled}.  The momentum $p$ is given by \eq{eq:momentum}. Note that all the numerical values for $\tilde{\sigma}$ reported in the paper are in excellent agreement with \eq{eq:sigma}, with a relative difference of the order of a percent.

Now, we compare our findings with those obtained in Ref.~\cite{Meinert945} at strong boson-boson and impurity-boson interaction for $M=m$. As for weakly-interacting bosons, the impurity drift velocity monotonically increases with $F$ in the Tonks-Girardeau gas. However, a monotonically decreasing Bloch amplitude with $F$ for the Tonks-Girardeau gas is reported, in a stark contrast to the results obtained in previous sections. Sinusoidal-like oscillations are found, and the time period becomes somewhat bigger with respect to $2\pi\hbar n_0/F$ at large $F$, similarly to our results for $\tilde{G}\gg 1$ at $\gamma \ll1$. The Bloch oscillations are found to be stable in a broad range of forces. Neither the nature of the emitted excitations, nor the impurity mobility was studied in Ref.~\cite{Meinert945}.
Moreover, the zero-temperature Bloch oscillations are strongly damped at strong boson-boson interaction \cite{PhysRevE.90.032132,Meinert945}. 
In the case of weakly-interacting bosons, the phase coherence length is huge, $\ell_{\phi}=\xi \exp{(2\sqrt{\pi^2/\gamma})}$ \cite{Petrov}. The effects of quantum fluctuations can be treated perturbatively \cite{sykes_drag_2009,CasimirNewJPhys}, and lead to a small damping of Bloch oscillations at small $\gamma$ \cite{Schecter_2016}. Increasing $\gamma$, the damping increases. 

A mobile impurity weakly-coupled to the Tonks-Girardeau gas and driven by a weak external force has been studied in Ref.~\cite{PhysRevE.90.032132}. It was predicted that the Bloch oscillations take place only if the impurity is heavier than the host particles. In the case of a lighter impurity, the drift velocity remains constant in time, while the velocity oscillations are absent. Also, it was found that at small $F$, the drift velocity scales as $\sqrt{F}$ for a heavy impurity, while it remains finite in the limit $F\to 0$ for a light impurity. In contrast, we do not find any dramatic change of the dynamics as a function of the impurity mass in the case of weakly-interacting bosons, as discussed above. Moreover, the Bloch amplitude decreases with the impurity mass at small force.

\section{Conclusions and discussion\label{sec:conclusions}}

In this work, we have studied the far-from-equilibrium dynamics of a system consisting of an impurity subject to an external force and an environment made of weakly-interacting 1d bosons at zero temperature. We have focused on the dynamics of both the impurity and the background bosons, revealing different dynamical regimes. We summarize below some of our main findings.
 
For an infinitesimal force, the system follows its finite-momentum ground state in time, which is a periodic function of momentum. Thus, the impurity velocity and the local depletion of the boson density situated at the impurity position oscillate periodically, while the momentum pumped into the system per time period is carried by the superfluid currents. Since the system is gapless, as the force is increased, the system moves away from its ground state by emitting more and more energetic excitations in the form of dispersive density shock waves, solitons and density waves. Contrary to the expectations \cite{AnnalsKamenev}, even-though the emission of nonlinear excitations takes place, it does not ruin the impurity Bloch oscillations in a wide interval of the driving force. The excitations are periodically emitted, and propagate away from the impurity.  At sufficiently big force, the Bloch oscillations cease and the impurity exhibits an unlimited acceleration.

Increasing the driving force $F$, the impurity drift velocity increases and moves away from the regime of linear dependence on $F$. The drift velocity exhibits a superlinaer and a sublinaer dependence on $F$ at weak-to-moderate and strong impurity coupling, respectively. The total growth of the phase variation along the system length over the period of oscillations is $2\pi$.
The Bloch amplitude $V_B$ displays a non-monotonic behaviour with $F$.  At small $F$, $V_B$ is an increasing function. At weak-to-moderate impurity coupling $\tilde{G}$, after reaching its maximal value, $V_B$ abruptly drops as $F$ is increased. Simultaneously, the drift velocity gets abruptly enhanced. At the value of $F$ where $V_B$ is maximal, the time period of oscillations is just a few percents shorter than the estimation $2\pi\hbar n_0/F$. However, for bigger forces where the Bloch oscillations still persist,  the period deviates strongly  from the estimated one and the underlying mechanism gets modified. The increase of the phase variation along the system length over the period becomes zero. 

In contrast to this scenario, at strong impurity coupling, after reaching its maximal value, $V_B$ smoothly decreases with $F$. This marks a crossover into regime where the deviation of the drift velocity from the linear $F$-dependence, as well as the deviation of the period from the aforementioned estimation, becomes important. 
The period is longer than the estimated one, while the increase of the phase variation along the system length over the period remains $2\pi$. Moreover, at a larger driving force, a different dynamics can be realised, where the amplitude of the oscillations becomes zero, while the drift velocity remains constant in time.

The maximal driving force where the Bloch mechanism exists, increases with the impurity-boson coupling, while the impurity drift velocity monotonically decreases. Moreover, the impurity velocity oscillations change in shape and become more sinusoidal-like with increasing $\tilde{G}$. The interval of $F$ where the drift velocity linearly depends on the force becomes wider at bigger $\tilde{G}$.

Increasing the impurity to boson mass ratio, while keeping the other parameters fixed, the drift velocity $V_d$ decreases. Furthermore, the interval of forces where $V_d$ exhibits the linear behavior with $F$ increases with $M/m$. If the boson-boson interaction strength is increased, the drift velocity diminishes, while the Bloch amplitude gets enhanced.

At low temperatures, the impurity experiences an additional friction due to scattering off thermally excited quasiparticles, leading to a small temperature correction of the obtained results \cite{AnnalsKamenev}. The thermal friction has been studied in the equilibrium conditions, i.e., for an impurity lying on the ground-state energy-momentum curve, and thus having a velocity smaller than the critical one \cite{castro_neto1996dynamics,gangardt2009bloch,PRLimpurity,PhysRevB.101.104503,MyPRLDynamics2023}. We have shown that the Bloch oscillations at zero temperature persist even though the impurity velocity is higher than both the critical and the sound velocity. The analysis of the thermal friction in the presence of a finite driving force is left for future studies. However, the obtained dependence of the zero-temperature properties of the Bloch oscillations on the system parameters indicates in which part of the parameter space the oscillations are expected to be the most robust.

\section*{Acknowledgments}

This study has been partially supported through the EUR grant NanoX n° ANR-17-EURE-0009 in the framework of the “Programme des Investissements d’Avenir”.



\begin{thebibliography}{35}%
	\makeatletter
	\providecommand \@ifxundefined [1]{%
		\@ifx{#1\undefined}
	}%
	\providecommand \@ifnum [1]{%
		\ifnum #1\expandafter \@firstoftwo
		\else \expandafter \@secondoftwo
		\fi
	}%
	\providecommand \@ifx [1]{%
		\ifx #1\expandafter \@firstoftwo
		\else \expandafter \@secondoftwo
		\fi
	}%
	\providecommand \natexlab [1]{#1}%
	\providecommand \enquote  [1]{``#1''}%
	\providecommand \bibnamefont  [1]{#1}%
	\providecommand \bibfnamefont [1]{#1}%
	\providecommand \citenamefont [1]{#1}%
	\providecommand \href@noop [0]{\@secondoftwo}%
	\providecommand \href [0]{\begingroup \@sanitize@url \@href}%
	\providecommand \@href[1]{\@@startlink{#1}\@@href}%
	\providecommand \@@href[1]{\endgroup#1\@@endlink}%
	\providecommand \@sanitize@url [0]{\catcode `\\12\catcode `\$12\catcode
		`\&12\catcode `\#12\catcode `\^12\catcode `\_12\catcode `\%12\relax}%
	\providecommand \@@startlink[1]{}%
	\providecommand \@@endlink[0]{}%
	\providecommand \url  [0]{\begingroup\@sanitize@url \@url }%
	\providecommand \@url [1]{\endgroup\@href {#1}{\urlprefix }}%
	\providecommand \urlprefix  [0]{URL }%
	\providecommand \Eprint [0]{\href }%
	\providecommand \doibase [0]{https://doi.org/}%
	\providecommand \selectlanguage [0]{\@gobble}%
	\providecommand \bibinfo  [0]{\@secondoftwo}%
	\providecommand \bibfield  [0]{\@secondoftwo}%
	\providecommand \translation [1]{[#1]}%
	\providecommand \BibitemOpen [0]{}%
	\providecommand \bibitemStop [0]{}%
	\providecommand \bibitemNoStop [0]{.\EOS\space}%
	\providecommand \EOS [0]{\spacefactor3000\relax}%
	\providecommand \BibitemShut  [1]{\csname bibitem#1\endcsname}%
	\let\auto@bib@innerbib\@empty
	\bibitem [{\citenamefont {Bloch}(1929)}]{Bloch}%
	\BibitemOpen
	\bibfield  {author} {\bibinfo {author} {\bibfnamefont {F.}~\bibnamefont
			{Bloch}},\ }\bibfield  {title} {\bibinfo {title} {{\"U}ber die
			quantenmechanik der elektronen in kristallgittern},\ }\href
	{https://doi.org/https://doi.org/10.1007/BF01339455} {\bibfield  {journal}
		{\bibinfo  {journal} {Zeitschrift f\"ur Physik}\ }\textbf {\bibinfo {volume}
			{52}},\ \bibinfo {pages} {555} (\bibinfo {year} {1929})}\BibitemShut
	{NoStop}%
	\bibitem [{\citenamefont {Zener}(1934)}]{Zener}%
	\BibitemOpen
	\bibfield  {author} {\bibinfo {author} {\bibfnamefont {C.}~\bibnamefont
			{Zener}},\ }\href {https://doi.org/http://doi.org/10.1098/rspa.1934.0116}
	{\bibfield  {journal} {\bibinfo  {journal} {Proc. R. Soc. Lond.}\ }\textbf
		{\bibinfo {volume} {145}},\ \bibinfo {pages} {523} (\bibinfo {year}
		{1934})}\BibitemShut {NoStop}%
	\bibitem [{\citenamefont {Waschke}\ \emph {et~al.}(1993)\citenamefont
		{Waschke}, \citenamefont {Roskos}, \citenamefont {Schwedler}, \citenamefont
		{Leo}, \citenamefont {Kurz},\ and\ \citenamefont
		{K\"ohler}}]{PhysRevLett.70.3319}%
	\BibitemOpen
	\bibfield  {author} {\bibinfo {author} {\bibfnamefont {C.}~\bibnamefont
			{Waschke}}, \bibinfo {author} {\bibfnamefont {H.~G.}\ \bibnamefont {Roskos}},
		\bibinfo {author} {\bibfnamefont {R.}~\bibnamefont {Schwedler}}, \bibinfo
		{author} {\bibfnamefont {K.}~\bibnamefont {Leo}}, \bibinfo {author}
		{\bibfnamefont {H.}~\bibnamefont {Kurz}},\ and\ \bibinfo {author}
		{\bibfnamefont {K.}~\bibnamefont {K\"ohler}},\ }\bibfield  {title} {\bibinfo
		{title} {Coherent submillimeter-wave emission from {Bloch} oscillations in a
			semiconductor superlattice},\ }\href
	{https://doi.org/10.1103/PhysRevLett.70.3319} {\bibfield  {journal} {\bibinfo
			{journal} {Phys. Rev. Lett.}\ }\textbf {\bibinfo {volume} {70}},\ \bibinfo
		{pages} {3319} (\bibinfo {year} {1993})}\BibitemShut {NoStop}%
	\bibitem [{\citenamefont {Ben~Dahan}\ \emph {et~al.}(1996)\citenamefont
		{Ben~Dahan}, \citenamefont {Peik}, \citenamefont {Reichel}, \citenamefont
		{Castin},\ and\ \citenamefont {Salomon}}]{PhysRevLett.76.4508}%
	\BibitemOpen
	\bibfield  {author} {\bibinfo {author} {\bibfnamefont {M.}~\bibnamefont
			{Ben~Dahan}}, \bibinfo {author} {\bibfnamefont {E.}~\bibnamefont {Peik}},
		\bibinfo {author} {\bibfnamefont {J.}~\bibnamefont {Reichel}}, \bibinfo
		{author} {\bibfnamefont {Y.}~\bibnamefont {Castin}},\ and\ \bibinfo {author}
		{\bibfnamefont {C.}~\bibnamefont {Salomon}},\ }\bibfield  {title} {\bibinfo
		{title} {Bloch oscillations of atoms in an optical potential},\ }\href
	{https://doi.org/10.1103/PhysRevLett.76.4508} {\bibfield  {journal} {\bibinfo
			{journal} {Phys. Rev. Lett.}\ }\textbf {\bibinfo {volume} {76}},\ \bibinfo
		{pages} {4508} (\bibinfo {year} {1996})}\BibitemShut {NoStop}%
	\bibitem [{\citenamefont {Wilkinson}\ \emph {et~al.}(1996)\citenamefont
		{Wilkinson}, \citenamefont {Bharucha}, \citenamefont {Madison}, \citenamefont
		{Niu},\ and\ \citenamefont {Raizen}}]{PhysRevLett.76.4512}%
	\BibitemOpen
	\bibfield  {author} {\bibinfo {author} {\bibfnamefont {S.~R.}\ \bibnamefont
			{Wilkinson}}, \bibinfo {author} {\bibfnamefont {C.~F.}\ \bibnamefont
			{Bharucha}}, \bibinfo {author} {\bibfnamefont {K.~W.}\ \bibnamefont
			{Madison}}, \bibinfo {author} {\bibfnamefont {Q.}~\bibnamefont {Niu}},\ and\
		\bibinfo {author} {\bibfnamefont {M.~G.}\ \bibnamefont {Raizen}},\ }\bibfield
	{title} {\bibinfo {title} {Observation of atomic {Wannier-Stark} ladders in
			an accelerating optical potential},\ }\href
	{https://doi.org/10.1103/PhysRevLett.76.4512} {\bibfield  {journal} {\bibinfo
			{journal} {Phys. Rev. Lett.}\ }\textbf {\bibinfo {volume} {76}},\ \bibinfo
		{pages} {4512} (\bibinfo {year} {1996})}\BibitemShut {NoStop}%
	\bibitem [{\citenamefont {Gangardt}\ and\ \citenamefont
		{Kamenev}(2009)}]{gangardt2009bloch}%
	\BibitemOpen
	\bibfield  {author} {\bibinfo {author} {\bibfnamefont {D.~M.}\ \bibnamefont
			{Gangardt}}\ and\ \bibinfo {author} {\bibfnamefont {A.}~\bibnamefont
			{Kamenev}},\ }\bibfield  {title} {\bibinfo {title} {Bloch {Oscillations} in a
			{One}-{Dimensional} {Spinor} {Gas}},\ }\href
	{https://doi.org/10.1103/PhysRevLett.102.070402} {\bibfield  {journal}
		{\bibinfo  {journal} {Phys. Rev. Lett.}\ }\textbf {\bibinfo {volume} {102}},\
		\bibinfo {pages} {070402} (\bibinfo {year} {2009})}\BibitemShut {NoStop}%
	\bibitem [{\citenamefont {Lamacraft}(2009)}]{lamacraft2009dispersion}%
	\BibitemOpen
	\bibfield  {author} {\bibinfo {author} {\bibfnamefont {A.}~\bibnamefont
			{Lamacraft}},\ }\bibfield  {title} {\bibinfo {title} {Dispersion relation and
			spectral function of an impurity in a one-dimensional quantum liquid},\
	}\href {https://doi.org/10.1103/PhysRevB.79.241105} {\bibfield  {journal}
		{\bibinfo  {journal} {Phys. Rev. B}\ }\textbf {\bibinfo {volume} {79}},\
		\bibinfo {pages} {241105} (\bibinfo {year} {2009})}\BibitemShut {NoStop}%
	\bibitem [{\citenamefont {Meinert}\ \emph {et~al.}(2017)\citenamefont
		{Meinert}, \citenamefont {Knap}, \citenamefont {Kirilov}, \citenamefont
		{Jag-Lauber}, \citenamefont {Zvonarev}, \citenamefont {Demler},\ and\
		\citenamefont {N{\"a}gerl}}]{Meinert945}%
	\BibitemOpen
	\bibfield  {author} {\bibinfo {author} {\bibfnamefont {F.}~\bibnamefont
			{Meinert}}, \bibinfo {author} {\bibfnamefont {M.}~\bibnamefont {Knap}},
		\bibinfo {author} {\bibfnamefont {E.}~\bibnamefont {Kirilov}}, \bibinfo
		{author} {\bibfnamefont {K.}~\bibnamefont {Jag-Lauber}}, \bibinfo {author}
		{\bibfnamefont {M.~B.}\ \bibnamefont {Zvonarev}}, \bibinfo {author}
		{\bibfnamefont {E.}~\bibnamefont {Demler}},\ and\ \bibinfo {author}
		{\bibfnamefont {H.-C.}\ \bibnamefont {N{\"a}gerl}},\ }\bibfield  {title}
	{\bibinfo {title} {Bloch oscillations in the absence of a lattice},\ }\href
	{https://doi.org/10.1126/science.aah6616} {\bibfield  {journal} {\bibinfo
			{journal} {Science}\ }\textbf {\bibinfo {volume} {356}},\ \bibinfo {pages}
		{945} (\bibinfo {year} {2017})}\BibitemShut {NoStop}%
	\bibitem [{\citenamefont {Kosevich}(2001)}]{ThBlochSolitons}%
	\BibitemOpen
	\bibfield  {author} {\bibinfo {author} {\bibfnamefont {A.~M.}\ \bibnamefont
			{Kosevich}},\ }\bibfield  {title} {\bibinfo {title} {Bloch oscillations of
			magnetic solitons as an example of dynamical localization of quasiparticles
			in a uniform external field},\ }\href {https://doi.org/10.1063/1.1388415}
	{\bibfield  {journal} {\bibinfo  {journal} {Low Temp. Phys.}\ }\textbf
		{\bibinfo {volume} {27}} (\bibinfo {year} {2001})}\BibitemShut {NoStop}%
	\bibitem [{\citenamefont {Bresolin}\ \emph {et~al.}(2023)\citenamefont
		{Bresolin}, \citenamefont {Roy}, \citenamefont {Ferrari}, \citenamefont
		{Recati},\ and\ \citenamefont {Pavloff}}]{PhysRevLett.130.220403}%
	\BibitemOpen
	\bibfield  {author} {\bibinfo {author} {\bibfnamefont {S.}~\bibnamefont
			{Bresolin}}, \bibinfo {author} {\bibfnamefont {A.}~\bibnamefont {Roy}},
		\bibinfo {author} {\bibfnamefont {G.}~\bibnamefont {Ferrari}}, \bibinfo
		{author} {\bibfnamefont {A.}~\bibnamefont {Recati}},\ and\ \bibinfo {author}
		{\bibfnamefont {N.}~\bibnamefont {Pavloff}},\ }\bibfield  {title} {\bibinfo
		{title} {Oscillating solitons and ac {Josephson} effect in ferromagnetic
			{Bose-Bose} mixtures},\ }\href
	{https://doi.org/10.1103/PhysRevLett.130.220403} {\bibfield  {journal}
		{\bibinfo  {journal} {Phys. Rev. Lett.}\ }\textbf {\bibinfo {volume} {130}},\
		\bibinfo {pages} {220403} (\bibinfo {year} {2023})}\BibitemShut {NoStop}%
	\bibitem [{\citenamefont {Rabec}\ \emph {et~al.}(2025)\citenamefont {Rabec},
		\citenamefont {Chauveau}, \citenamefont {Brochier}, \citenamefont
		{Nascimbene}, \citenamefont {Dalibard},\ and\ \citenamefont
		{Beugnon}}]{BlochSolitons}%
	\BibitemOpen
	\bibfield  {author} {\bibinfo {author} {\bibfnamefont {F.}~\bibnamefont
			{Rabec}}, \bibinfo {author} {\bibfnamefont {G.}~\bibnamefont {Chauveau}},
		\bibinfo {author} {\bibfnamefont {G.}~\bibnamefont {Brochier}}, \bibinfo
		{author} {\bibfnamefont {S.}~\bibnamefont {Nascimbene}}, \bibinfo {author}
		{\bibfnamefont {J.}~\bibnamefont {Dalibard}},\ and\ \bibinfo {author}
		{\bibfnamefont {J.}~\bibnamefont {Beugnon}},\ }\bibfield  {title} {\bibinfo
		{title} {Bloch oscillations of a soliton in a 1d quantum fluid},\ }\href
	{https://doi.org/10.1038/s41567-025-02970-1} {\bibfield  {journal} {\bibinfo
			{journal} {Nature Physics}\ }\textbf {\bibinfo {volume} {21}},\ \bibinfo
		{pages} {1541} (\bibinfo {year} {2025})}\BibitemShut {NoStop}%
	\bibitem [{\citenamefont {Schecter}\ \emph
		{et~al.}(2012{\natexlab{a}})\citenamefont {Schecter}, \citenamefont
		{Gangardt},\ and\ \citenamefont {Kamenev}}]{AnnalsKamenev}%
	\BibitemOpen
	\bibfield  {author} {\bibinfo {author} {\bibfnamefont {M.}~\bibnamefont
			{Schecter}}, \bibinfo {author} {\bibfnamefont {D.}~\bibnamefont {Gangardt}},\
		and\ \bibinfo {author} {\bibfnamefont {A.}~\bibnamefont {Kamenev}},\
	}\bibfield  {title} {\bibinfo {title} {Dynamics and {Bloch} oscillations of
			mobile impurities in one-dimensional quantum liquids},\ }\href
	{https://doi.org/https://doi.org/10.1016/j.aop.2011.10.001} {\bibfield
		{journal} {\bibinfo  {journal} {Ann. Phys.}\ }\textbf {\bibinfo {volume}
			{327}},\ \bibinfo {pages} {639} (\bibinfo {year}
		{2012}{\natexlab{a}})}\BibitemShut {NoStop}%
	\bibitem [{\citenamefont {Schecter}\ \emph {et~al.}(2016)\citenamefont
		{Schecter}, \citenamefont {Gangardt},\ and\ \citenamefont
		{Kamenev}}]{Schecter_2016}%
	\BibitemOpen
	\bibfield  {author} {\bibinfo {author} {\bibfnamefont {M.}~\bibnamefont
			{Schecter}}, \bibinfo {author} {\bibfnamefont {D.~M.}\ \bibnamefont
			{Gangardt}},\ and\ \bibinfo {author} {\bibfnamefont {A.}~\bibnamefont
			{Kamenev}},\ }\bibfield  {title} {\bibinfo {title} {Quantum impurities: from
			mobile {Josephson} junctions to depletons},\ }\href
	{https://doi.org/10.1088/1367-2630/18/6/065002} {\bibfield  {journal}
		{\bibinfo  {journal} {New J. Phys.}\ }\textbf {\bibinfo {volume} {18}},\
		\bibinfo {pages} {065002} (\bibinfo {year} {2016})}\BibitemShut {NoStop}%
	\bibitem [{\citenamefont {Gamayun}\ \emph {et~al.}(2014)\citenamefont
		{Gamayun}, \citenamefont {Lychkovskiy},\ and\ \citenamefont
		{Cheianov}}]{PhysRevE.90.032132}%
	\BibitemOpen
	\bibfield  {author} {\bibinfo {author} {\bibfnamefont {O.}~\bibnamefont
			{Gamayun}}, \bibinfo {author} {\bibfnamefont {O.}~\bibnamefont
			{Lychkovskiy}},\ and\ \bibinfo {author} {\bibfnamefont {V.}~\bibnamefont
			{Cheianov}},\ }\bibfield  {title} {\bibinfo {title} {Kinetic theory for a
			mobile impurity in a degenerate {Tonks}-{Girardeau} gas},\ }\href
	{https://doi.org/10.1103/PhysRevE.90.032132} {\bibfield  {journal} {\bibinfo
			{journal} {Phys. Rev. E}\ }\textbf {\bibinfo {volume} {90}},\ \bibinfo
		{pages} {032132} (\bibinfo {year} {2014})}\BibitemShut {NoStop}%
	\bibitem [{\citenamefont {Schecter}\ \emph {et~al.}(2015)\citenamefont
		{Schecter}, \citenamefont {Gangardt},\ and\ \citenamefont
		{Kamenev}}]{PhysRevE.92.016101}%
	\BibitemOpen
	\bibfield  {author} {\bibinfo {author} {\bibfnamefont {M.}~\bibnamefont
			{Schecter}}, \bibinfo {author} {\bibfnamefont {D.~M.}\ \bibnamefont
			{Gangardt}},\ and\ \bibinfo {author} {\bibfnamefont {A.}~\bibnamefont
			{Kamenev}},\ }\bibfield  {title} {\bibinfo {title} {Comment on ``{Kinetic}
			theory for a mobile impurity in a degenerate {Tonks-Girardeau} gas''},\
	}\href {https://doi.org/10.1103/PhysRevE.92.016101} {\bibfield  {journal}
		{\bibinfo  {journal} {Phys. Rev. E}\ }\textbf {\bibinfo {volume} {92}},\
		\bibinfo {pages} {016101} (\bibinfo {year} {2015})}\BibitemShut {NoStop}%
	\bibitem [{\citenamefont {Gamayun}\ \emph {et~al.}(2015)\citenamefont
		{Gamayun}, \citenamefont {Lychkovskiy},\ and\ \citenamefont
		{Cheianov}}]{PhysRevE.92.016102}%
	\BibitemOpen
	\bibfield  {author} {\bibinfo {author} {\bibfnamefont {O.}~\bibnamefont
			{Gamayun}}, \bibinfo {author} {\bibfnamefont {O.}~\bibnamefont
			{Lychkovskiy}},\ and\ \bibinfo {author} {\bibfnamefont {V.}~\bibnamefont
			{Cheianov}},\ }\bibfield  {title} {\bibinfo {title} {Reply to ``{Comment} on
			`{Kinetic} theory for a mobile impurity in a degenerate {Tonks-Girardeau}
			gas' ''},\ }\href {https://doi.org/10.1103/PhysRevE.92.016102} {\bibfield
		{journal} {\bibinfo  {journal} {Phys. Rev. E}\ }\textbf {\bibinfo {volume}
			{92}},\ \bibinfo {pages} {016102} (\bibinfo {year} {2015})}\BibitemShut
	{NoStop}%
	\bibitem [{\citenamefont {Lee}\ \emph {et~al.}(1953)\citenamefont {Lee},
		\citenamefont {Low},\ and\ \citenamefont {Pines}}]{LeeLowPines}%
	\BibitemOpen
	\bibfield  {author} {\bibinfo {author} {\bibfnamefont {T.~D.}\ \bibnamefont
			{Lee}}, \bibinfo {author} {\bibfnamefont {F.~E.}\ \bibnamefont {Low}},\ and\
		\bibinfo {author} {\bibfnamefont {D.}~\bibnamefont {Pines}},\ }\bibfield
	{title} {\bibinfo {title} {The motion of slow electrons in a polar crystal},\
	}\href {https://doi.org/https://doi.org/10.1103/PhysRev.90.297} {\bibfield
		{journal} {\bibinfo  {journal} {Phys. Rev.}\ }\textbf {\bibinfo {volume}
			{90}},\ \bibinfo {pages} {297} (\bibinfo {year} {1953})}\BibitemShut
	{NoStop}%
	\bibitem [{\citenamefont {Pitaevskii}\ and\ \citenamefont
		{Stringari}(2003)}]{pitaevskii_bose-einstein_2003}%
	\BibitemOpen
	\bibfield  {author} {\bibinfo {author} {\bibfnamefont {L.~P.}\ \bibnamefont
			{Pitaevskii}}\ and\ \bibinfo {author} {\bibfnamefont {S.}~\bibnamefont
			{Stringari}},\ }\href@noop {} {\emph {\bibinfo {title} {Bose-{Einstein}
				{Condensation}}}},\ International {Series} of {Monographs} on {Physics}\
	(\bibinfo  {publisher} {Oxford University Press},\ \bibinfo {address}
	{Oxford, New York},\ \bibinfo {year} {2003})\BibitemShut {NoStop}%
	\bibitem [{\citenamefont {Sykes}\ \emph {et~al.}(2009)\citenamefont {Sykes},
		\citenamefont {Davis},\ and\ \citenamefont {Roberts}}]{sykes_drag_2009}%
	\BibitemOpen
	\bibfield  {author} {\bibinfo {author} {\bibfnamefont {A.~G.}\ \bibnamefont
			{Sykes}}, \bibinfo {author} {\bibfnamefont {M.~J.}\ \bibnamefont {Davis}},\
		and\ \bibinfo {author} {\bibfnamefont {D.~C.}\ \bibnamefont {Roberts}},\
	}\bibfield  {title} {\bibinfo {title} {Drag force on an impurity below the
			superfluid critical velocity in a quasi-one-dimensional {Bose}-{Einstein}
			condensate},\ }\href {https://doi.org/10.1103/PhysRevLett.103.085302}
	{\bibfield  {journal} {\bibinfo  {journal} {Phys. Rev. Lett.}\ }\textbf
		{\bibinfo {volume} {103}},\ \bibinfo {pages} {085302} (\bibinfo {year}
		{2009})}\BibitemShut {NoStop}%
	\bibitem [{\citenamefont {Reichert}\ \emph {et~al.}(2019)\citenamefont
		{Reichert}, \citenamefont {Ristivojevic},\ and\ \citenamefont
		{Petkovi{\'{c}}}}]{CasimirNewJPhys}%
	\BibitemOpen
	\bibfield  {author} {\bibinfo {author} {\bibfnamefont {B.}~\bibnamefont
			{Reichert}}, \bibinfo {author} {\bibfnamefont {Z.}~\bibnamefont
			{Ristivojevic}},\ and\ \bibinfo {author} {\bibfnamefont {A.}~\bibnamefont
			{Petkovi{\'{c}}}},\ }\bibfield  {title} {\bibinfo {title} {The {Casimir-like}
			effect in a one-dimensional {Bose} gas},\ }\href
	{https://doi.org/10.1088/1367-2630/ab1b8e} {\bibfield  {journal} {\bibinfo
			{journal} {New J. Phys.}\ }\textbf {\bibinfo {volume} {21}},\ \bibinfo
		{pages} {053024} (\bibinfo {year} {2019})}\BibitemShut {NoStop}%
	\bibitem [{\citenamefont {Mistakidis}\ \emph {et~al.}(2019)\citenamefont
		{Mistakidis}, \citenamefont {Volosniev}, \citenamefont {Zinner},\ and\
		\citenamefont {Schmelcher}}]{PhysRevA.100.013619}%
	\BibitemOpen
	\bibfield  {author} {\bibinfo {author} {\bibfnamefont {S.~I.}\ \bibnamefont
			{Mistakidis}}, \bibinfo {author} {\bibfnamefont {A.~G.}\ \bibnamefont
			{Volosniev}}, \bibinfo {author} {\bibfnamefont {N.~T.}\ \bibnamefont
			{Zinner}},\ and\ \bibinfo {author} {\bibfnamefont {P.}~\bibnamefont
			{Schmelcher}},\ }\bibfield  {title} {\bibinfo {title} {Effective approach to
			impurity dynamics in one-dimensional trapped {Bose} gases},\ }\href
	{https://doi.org/10.1103/PhysRevA.100.013619} {\bibfield  {journal} {\bibinfo
			{journal} {Phys. Rev. A}\ }\textbf {\bibinfo {volume} {100}},\ \bibinfo
		{pages} {013619} (\bibinfo {year} {2019})}\BibitemShut {NoStop}%
	\bibitem [{\citenamefont {Koutentakis}\ \emph {et~al.}(2022)\citenamefont
		{Koutentakis}, \citenamefont {Mistakidis},\ and\ \citenamefont
		{Schmelcher}}]{atoms10010003}%
	\BibitemOpen
	\bibfield  {author} {\bibinfo {author} {\bibfnamefont {G.~M.}\ \bibnamefont
			{Koutentakis}}, \bibinfo {author} {\bibfnamefont {S.~I.}\ \bibnamefont
			{Mistakidis}},\ and\ \bibinfo {author} {\bibfnamefont {P.}~\bibnamefont
			{Schmelcher}},\ }\bibfield  {title} {\bibinfo {title} {Pattern formation in
			one-dimensional polaron systems and temporal orthogonality catastrophe},\
	}\href {https://doi.org/10.3390/atoms10010003} {\bibfield  {journal}
		{\bibinfo  {journal} {Atoms}\ }\textbf {\bibinfo {volume} {10}},\ \bibinfo
		{pages} {3} (\bibinfo {year} {2022})}\BibitemShut {NoStop}%
	\bibitem [{\citenamefont {Will}\ and\ \citenamefont
		{Fleischhauer}(2023)}]{Will_2023}%
	\BibitemOpen
	\bibfield  {author} {\bibinfo {author} {\bibfnamefont {M.}~\bibnamefont
			{Will}}\ and\ \bibinfo {author} {\bibfnamefont {M.}~\bibnamefont
			{Fleischhauer}},\ }\bibfield  {title} {\bibinfo {title} {Dynamics of polaron
			formation in 1d {Bose} gases in the strong-coupling regime},\ }\href
	{https://doi.org/10.1088/1367-2630/acf06a} {\bibfield  {journal} {\bibinfo
			{journal} {New J. of Phys.}\ }\textbf {\bibinfo {volume} {25}},\ \bibinfo
		{pages} {083043} (\bibinfo {year} {2023})}\BibitemShut {NoStop}%
	\bibitem [{\citenamefont {Hakim}(1997)}]{Hakim}%
	\BibitemOpen
	\bibfield  {author} {\bibinfo {author} {\bibfnamefont {V.}~\bibnamefont
			{Hakim}},\ }\bibfield  {title} {\bibinfo {title} {Nonlinear {Schr\"odinger}
			flow past an obstacle in one dimension},\ }\href
	{https://doi.org/10.1103/PhysRevE.55.2835} {\bibfield  {journal} {\bibinfo
			{journal} {Phys. Rev. E}\ }\textbf {\bibinfo {volume} {55}},\ \bibinfo
		{pages} {2835} (\bibinfo {year} {1997})}\BibitemShut {NoStop}%
	\bibitem [{\citenamefont {Majumdar}\ and\ \citenamefont
		{Petkovi{\'{c}}}(2025)}]{Quench}%
	\BibitemOpen
	\bibfield  {author} {\bibinfo {author} {\bibfnamefont {S.}~\bibnamefont
			{Majumdar}}\ and\ \bibinfo {author} {\bibfnamefont {A.}~\bibnamefont
			{Petkovi{\'{c}}}},\ }\bibfield  {title} {\bibinfo {title} {Relaxation
			dynamics of a mobile impurity injected in a one-dimensional {Bose} gas},\
	}\href {https://doi.org/https://arxiv.org/abs/2507.10402} {\bibfield
		{journal} {\bibinfo  {journal} {arXiv:2507.10402}\ ,\ \bibinfo {pages} {to
				appear in Phys. Rev. A}} (\bibinfo {year} {2025})}\BibitemShut {NoStop}%
	\bibitem [{\citenamefont {Kamenev}\ and\ \citenamefont
		{Glazman}(2009)}]{kamenev2009dynamics}%
	\BibitemOpen
	\bibfield  {author} {\bibinfo {author} {\bibfnamefont {A.}~\bibnamefont
			{Kamenev}}\ and\ \bibinfo {author} {\bibfnamefont {L.~I.}\ \bibnamefont
			{Glazman}},\ }\bibfield  {title} {\bibinfo {title} {Dynamics of a
			one-dimensional spinor {Bose} liquid: {A} phenomenological approach},\ }\href
	{https://doi.org/10.1103/PhysRevA.80.011603} {\bibfield  {journal} {\bibinfo
			{journal} {Phys. Rev. A}\ }\textbf {\bibinfo {volume} {80}},\ \bibinfo
		{pages} {011603} (\bibinfo {year} {2009})}\BibitemShut {NoStop}%
	\bibitem [{\citenamefont {Trofimov}\ and\ \citenamefont
		{Peskov}(2009)}]{GPE_discretization}%
	\BibitemOpen
	\bibfield  {author} {\bibinfo {author} {\bibfnamefont {V.~A.}\ \bibnamefont
			{Trofimov}}\ and\ \bibinfo {author} {\bibfnamefont {N.~V.}\ \bibnamefont
			{Peskov}},\ }\bibfield  {title} {\bibinfo {title} {Comparison of
			finite difference schemes for the {Gross Pitaevskii} equation},\ }\href
	{https://doi.org/10.3846/1392-6292.2009.14.109-126} {\bibfield  {journal}
		{\bibinfo  {journal} {Mathematical Modelling and Analysis}\ }\textbf
		{\bibinfo {volume} {14}},\ \bibinfo {pages} {109} (\bibinfo {year}
		{2009})}\BibitemShut {NoStop}%
	\bibitem [{\citenamefont {Patankar}(1980)}]{upwind_scheme}%
	\BibitemOpen
	\bibfield  {author} {\bibinfo {author} {\bibfnamefont {S.}~\bibnamefont
			{Patankar}},\ }\href {https://books.google.fr/books?id=5JMYZMX3OVcC} {\emph
		{\bibinfo {title} {Numerical Heat Transfer and Fluid Flow}}},\ Series in
	computational methods in mechanics and thermal sciences\ (\bibinfo
	{publisher} {Taylor \& Francis},\ \bibinfo {year} {1980})\BibitemShut
	{NoStop}%
	\bibitem [{Note1()}]{Note1}%
	\BibitemOpen
	\bibinfo {note} {For the definition of the critical mass see Refs.~\cite
		{PhysRevLett.108.207001} and \cite {Quench}.}\BibitemShut {Stop}%
	\bibitem [{\citenamefont {{D.S. Petrov}}\ \emph {et~al.}(2004)\citenamefont
		{{D.S. Petrov}}, \citenamefont {{D.M. Gangardt}},\ and\ \citenamefont {{G.V.
				Shlyapnikov}}}]{Petrov}%
	\BibitemOpen
	\bibfield  {author} {\bibinfo {author} {\bibnamefont {{D.S. Petrov}}},
		\bibinfo {author} {\bibnamefont {{D.M. Gangardt}}},\ and\ \bibinfo {author}
		{\bibnamefont {{G.V. Shlyapnikov}}},\ }\bibfield  {title} {\bibinfo {title}
		{Low-dimensional trapped gases},\ }\href
	{https://doi.org/10.1051/jp4:2004116001} {\bibfield  {journal} {\bibinfo
			{journal} {J. Phys. IV France}\ }\textbf {\bibinfo {volume} {116}},\ \bibinfo
		{pages} {5} (\bibinfo {year} {2004})}\BibitemShut {NoStop}%
	\bibitem [{\citenamefont {Castro~Neto}\ and\ \citenamefont
		{Fisher}(1996)}]{castro_neto1996dynamics}%
	\BibitemOpen
	\bibfield  {author} {\bibinfo {author} {\bibfnamefont {A.~H.}\ \bibnamefont
			{Castro~Neto}}\ and\ \bibinfo {author} {\bibfnamefont {M.~P.~A.}\
			\bibnamefont {Fisher}},\ }\bibfield  {title} {\bibinfo {title} {Dynamics of a
			heavy particle in a {Luttinger} liquid},\ }\href
	{https://doi.org/10.1103/PhysRevB.53.9713} {\bibfield  {journal} {\bibinfo
			{journal} {Phys. Rev. B}\ }\textbf {\bibinfo {volume} {53}},\ \bibinfo
		{pages} {9713} (\bibinfo {year} {1996})}\BibitemShut {NoStop}%
	\bibitem [{\citenamefont {Petkovi\ifmmode~\acute{c}\else \'{c}\fi{}}\ and\
		\citenamefont {Ristivojevic}(2016)}]{PRLimpurity}%
	\BibitemOpen
	\bibfield  {author} {\bibinfo {author} {\bibfnamefont {A.}~\bibnamefont
			{Petkovi\ifmmode~\acute{c}\else \'{c}\fi{}}}\ and\ \bibinfo {author}
		{\bibfnamefont {Z.}~\bibnamefont {Ristivojevic}},\ }\bibfield  {title}
	{\bibinfo {title} {Dynamics of a mobile impurity in a one-dimensional {Bose}
			liquid},\ }\href {https://doi.org/10.1103/PhysRevLett.117.105301} {\bibfield
		{journal} {\bibinfo  {journal} {Phys. Rev. Lett.}\ }\textbf {\bibinfo
			{volume} {117}},\ \bibinfo {pages} {105301} (\bibinfo {year}
		{2016})}\BibitemShut {NoStop}%
	\bibitem [{\citenamefont {Petkovi\ifmmode~\acute{c}\else
			\'{c}\fi{}}(2020)}]{PhysRevB.101.104503}%
	\BibitemOpen
	\bibfield  {author} {\bibinfo {author} {\bibfnamefont {A.}~\bibnamefont
			{Petkovi\ifmmode~\acute{c}\else \'{c}\fi{}}},\ }\bibfield  {title} {\bibinfo
		{title} {Microscopic theory of the friction force exerted on a quantum
			impurity in one-dimensional quantum liquids},\ }\href
	{https://doi.org/10.1103/PhysRevB.101.104503} {\bibfield  {journal} {\bibinfo
			{journal} {Phys. Rev. B}\ }\textbf {\bibinfo {volume} {101}},\ \bibinfo
		{pages} {104503} (\bibinfo {year} {2020})}\BibitemShut {NoStop}%
	\bibitem [{\citenamefont {Petkovi\ifmmode~\acute{c}\else \'{c}\fi{}}\ and\
		\citenamefont {Ristivojevic}(2023)}]{MyPRLDynamics2023}%
	\BibitemOpen
	\bibfield  {author} {\bibinfo {author} {\bibfnamefont {A.}~\bibnamefont
			{Petkovi\ifmmode~\acute{c}\else \'{c}\fi{}}}\ and\ \bibinfo {author}
		{\bibfnamefont {Z.}~\bibnamefont {Ristivojevic}},\ }\bibfield  {title}
	{\bibinfo {title} {Dissipative dynamics of a heavy impurity in a {Bose} gas
			in the strong coupling regime},\ }\href
	{https://doi.org/10.1103/PhysRevLett.131.186001} {\bibfield  {journal}
		{\bibinfo  {journal} {Phys. Rev. Lett.}\ }\textbf {\bibinfo {volume} {131}},\
		\bibinfo {pages} {186001} (\bibinfo {year} {2023})}\BibitemShut {NoStop}%
	\bibitem [{\citenamefont {Schecter}\ \emph
		{et~al.}(2012{\natexlab{b}})\citenamefont {Schecter}, \citenamefont
		{Kamenev}, \citenamefont {Gangardt},\ and\ \citenamefont
		{Lamacraft}}]{PhysRevLett.108.207001}%
	\BibitemOpen
	\bibfield  {author} {\bibinfo {author} {\bibfnamefont {M.}~\bibnamefont
			{Schecter}}, \bibinfo {author} {\bibfnamefont {A.}~\bibnamefont {Kamenev}},
		\bibinfo {author} {\bibfnamefont {D.~M.}\ \bibnamefont {Gangardt}},\ and\
		\bibinfo {author} {\bibfnamefont {A.}~\bibnamefont {Lamacraft}},\ }\bibfield
	{title} {\bibinfo {title} {Critical velocity of a mobile impurity in
			one-dimensional quantum liquids},\ }\href
	{https://doi.org/10.1103/PhysRevLett.108.207001} {\bibfield  {journal}
		{\bibinfo  {journal} {Phys. Rev. Lett.}\ }\textbf {\bibinfo {volume} {108}},\
		\bibinfo {pages} {207001} (\bibinfo {year} {2012}{\natexlab{b}})}\BibitemShut
	{NoStop}%
\end{thebibliography}
%

\end{document}